\pdfoutput=1

\documentclass[aps,prd,preprint,superscriptaddress,preprintnumbers,nofootinbib,tightenlines,floatfix]{revtex4-1}

\usepackage{graphicx} 
\usepackage{dcolumn}  
\usepackage{bm}       

\begin{document}

\preprint{CLNS 12/2083}  
\preprint{CLEO 12-01}    

\title{\boldmath Updated Measurement of the Strong Phase in $D^0 \to K^+\pi^-$
Decay Using Quantum Correlations in $e^+e^- \to D^0 \bar D^0$ at CLEO}

\author{D.~M.~Asner}
\author{G.~Tatishvili}
\affiliation{Pacific Northwest National Laboratory, Richland, WA 99352, USA}
\author{J.~Y.~Ge}
\author{D.~H.~Miller}
\author{I.~P.~J.~Shipsey}
\author{B.~Xin}
\affiliation{Purdue University, West Lafayette, Indiana 47907, USA}
\author{G.~S.~Adams}
\author{J.~Napolitano}
\affiliation{Rensselaer Polytechnic Institute, Troy, New York 12180, USA}
\author{K.~M.~Ecklund}
\affiliation{Rice University, Houston, Texas 77005, USA}
\author{Q.~He}
\author{J.~Insler}
\author{H.~Muramatsu}
\author{L.~J.~Pearson}
\author{E.~H.~Thorndike}
\affiliation{University of Rochester, Rochester, New York 14627, USA}
\author{M.~Artuso}
\author{S.~Blusk}
\author{N.~Horwitz}
\author{R.~Mountain}
\author{T.~Skwarnicki}
\author{S.~Stone}
\author{J.~C.~Wang}
\author{L.~M.~Zhang}
\affiliation{Syracuse University, Syracuse, New York 13244, USA}
\author{P.~U.~E.~Onyisi}
\affiliation{University of Texas at Austin, Austin, TX 78712}
\author{G.~Bonvicini}
\author{D.~Cinabro}
\author{A.~Lincoln}
\author{M.~J.~Smith}
\author{P.~Zhou}
\affiliation{Wayne State University, Detroit, Michigan 48202, USA}
\author{P.~Naik}
\author{J.~Rademacker}\affiliation{University of Bristol, Bristol BS8 1TL, United Kingdom}
\author{K.~W.~Edwards}
\author{E.~J.~White}
\altaffiliation[Now at: ]{University of Cincinnati, Cincinnati, Ohio 45221}
\affiliation{Carleton University, Ottawa, Ontario, Canada K1S 5B6}
\author{R.~A.~Briere}
\author{H.~Vogel}
\affiliation{Carnegie Mellon University, Pittsburgh, Pennsylvania 15213, USA}
\author{P.~U.~E.~Onyisi}
\author{J.~L.~Rosner}
\affiliation{University of Chicago, Chicago, Illinois 60637, USA}
\author{J.~P.~Alexander}
\author{D.~G.~Cassel}
\author{S.~Das}
\author{R.~Ehrlich}
\author{L.~Gibbons}
\author{S.~W.~Gray}
\author{D.~L.~Hartill}
\author{D.~L.~Kreinick}
\author{V.~E.~Kuznetsov}
\author{J.~R.~Patterson}
\author{D.~Peterson}
\author{D.~Riley}
\author{A.~Ryd}
\author{A.~J.~Sadoff}
\author{X.~Shi}
\altaffiliation[Now at: ]{National Taiwan University, Taipei, Taiwan}
\author{S.~Stroiney}
\author{W.~M.~Sun}
\affiliation{Cornell University, Ithaca, New York 14853, USA}
\author{J.~Yelton}
\affiliation{University of Florida, Gainesville, Florida 32611, USA}
\author{P.~Rubin}
\affiliation{George Mason University, Fairfax, Virginia 22030, USA}
\author{N.~Lowrey}
\author{S.~Mehrabyan}
\author{M.~Selen}
\author{J.~Wiss}
\affiliation{University of Illinois, Urbana-Champaign, Illinois 61801, USA}
\author{J.~Libby}
\affiliation{Indian Institute of Technology Madras, Chennai, Tamil Nadu 600036, India}
\author{M.~Kornicer}
\author{R.~E.~Mitchell}
\affiliation{Indiana University, Bloomington, Indiana 47405, USA }
\author{D.~Besson}
\affiliation{University of Kansas, Lawrence, Kansas 66045, USA}
\author{T.~K.~Pedlar}
\affiliation{Luther College, Decorah, Iowa 52101, USA}
\author{D.~Cronin-Hennessy}
\author{J.~Hietala}
\affiliation{University of Minnesota, Minneapolis, Minnesota 55455, USA}
\author{S.~Dobbs}
\author{Z.~Metreveli}
\author{K.~K.~Seth}
\author{A.~Tomaradze}
\author{T.~Xiao}
\affiliation{Northwestern University, Evanston, Illinois 60208, USA}
\author{A.~Powell}
\author{C.~Thomas}
\author{G.~Wilkinson}
\affiliation{University of Oxford, Oxford OX1 3RH, United Kingdom}
\collaboration{CLEO Collaboration}
\noaffiliation

\date{October 2, 2012}

\begin{abstract} 
We analyze a sample of 3 million quantum-correlated $D^0\bar D^0$ pairs
from 818 ${\rm pb}^{-1}$ of $e^+e^-$ collision data collected with the
CLEO-c detector at $E_{\rm cm}=3.77$ GeV, to give an
updated measurement of $\cos\delta$ and a first determination of
$\sin\delta$, where $\delta$ is the relative strong phase between
doubly Cabibbo-suppressed $D^0\to K^+\pi^-$
and Cabibbo-favored $\bar D^0\to K^+\pi^-$ decay amplitudes.
With no inputs from other experiments, we find
$\cos\delta = 0.81^{+0.22+0.07}_{-0.18-0.05}$,
$\sin\delta = -0.01\pm 0.41 \pm 0.04$, and
$|\delta| = (10^{+28+13}_{-53-0})^\circ$.
By including external measurements of mixing parameters, we find
alternative values of $\cos\delta = 1.15^{+0.19+0.00}_{-0.17-0.08}$,
$\sin\delta = 0.56^{+0.32+0.21}_{-0.31-0.20}$,
and $\delta = (18^{+11}_{-17})^\circ$. Our results can be used
to improve the world average uncertainty on the mixing parameter
$y$ by approximately 10\%.
\end{abstract}

\pacs{12.15.Ff,13.20.Fc,13.25.Ft,14.40.Lb}
\maketitle

\section{Introduction}

Charm mixing in the Standard Model is conventionally described by two small
dimensionless parameters:
\begin{eqnarray}
x &\equiv& 2\frac{M_2 - M_1}{\Gamma_2 + \Gamma_1} \\
\label{eq:y}
y &\equiv& \frac{\Gamma_2 - \Gamma_1}{\Gamma_2 + \Gamma_1},
\end{eqnarray}
where $M_{1,2}$ and $\Gamma_{1,2}$ are the masses and widths, respectively,
of the neutral $D$ meson $CP$ eigenstates, $D_1$ ($CP$-odd) and
$D_2$ ($CP$-even), defined by
\begin{eqnarray}
\label{eq:d1}
|D_1\rangle \equiv \frac{|D^0\rangle + |\bar D^0\rangle}{\sqrt{2}} \\
\label{eq:d2}
|D_2\rangle \equiv \frac{|D^0\rangle - |\bar D^0\rangle}{\sqrt{2}},
\end{eqnarray}
assuming $CP$ conservation.  The mixing probability is then denoted by
$R_{\rm M}\equiv (x^2+y^2)/2$, and the width of the $D^0$ and $\bar D^0$
flavor eigenstates is $\Gamma\equiv (\Gamma_1+\Gamma_2)/2$.

Recent experimental studies of charm mixing parameters have probed $x$ and $y$
directly~\cite{ycpBelle1,ycpBABAR1,ycpBABAR2,ycpBelle2,ycpLHCb,kspipiBelle,kspipiBABAR},
as well as the ``rotated'' parameter
$y'\equiv y\cos\delta - x\sin\delta$~\cite{kpiBelle,kpiBABAR,kpiCDF}.
Here, $-\delta$ is the relative phase between the doubly
Cabibbo-suppressed $D^0\to K^+\pi^-$ amplitude and the corresponding Cabibbo-favored
$\bar D^0\to K^+\pi^-$ amplitude:
$\langle K^+\pi^-|D^0\rangle / \langle K^+\pi^-|\bar D^0\rangle\equiv r e^{-i\delta}$.
We adopt a convention in which $\delta$ corresponds to a strong
phase, which vanishes in the SU(3) limit~\cite{Gronau:2001nr}.
The magnitude $r$ of the amplitude ratio is approximately 0.06.
In this article, we update an analysis~\cite{tqca1} that first directly
determined
$\cos\delta$ using correlated production of $D^0$ and $\bar D^0$ mesons 
in $e^+e^-$ collisions produced at the Cornell Electron Storage Ring and
collected with the CLEO-c detector.
In the current analysis, we also present a first measurement of
$\sin\delta$.

At the $\psi(3770)$ resonance, the $D^0\bar D^0$ pair is produced with
no accompanying particles, so it is in a quantum-coherent $C=-1$ state.
As a result, the $D^0\bar D^0$ decay rates differ from incoherent
decay rates because of interference effects.
These differences depend on $y$ (to first order) and on
strong phases of the decay amplitudes~\cite{Kingsley:1975fe,Okun:1975di,Kingsley:1976cd,Goldhaber:1976fp,Bigi:1986dp,Bigi:1986rj,Bigi:1989ah,Xing:1996pn,Gronau:2001nr,Bianco:2003vb,Atwood:2002ak}.
As in our previous analysis, we implement the double-tagging
method with a $\chi^2$ fit described in Ref.~\cite{Asner:2005wf}, where,
in addition to extracting the number of $D^0\bar D^0$ pairs produced
(${\cal N}$) and the branching fractions (${\cal B}$) of the reconstructed
$D^0$ final states, we simultaneously determine
$y$, $x^2$, $r^2$, $\cos\delta$, and $\sin\delta$,
all without needing to know the integrated luminosity or $D^0\bar D^0$
production cross section.
The main improvements in the current analysis are:
use of the full CLEO-c $\psi(3770)$ dataset (which is three times larger than
the previous dataset), reconstruction of additional $CP$ eigenstates,
reconstruction of semimuonic $D^0$ decays, addition of modes that provide
sensitivity to $\sin\delta$, and direct measurement of the amplitude
ratio $r$.

As before, we neglect $CP$ violation in $D$ decays and mixing.
Recently, evidence has been found~\cite{LHCbCPV, CDFCPV1, CDFCPV2}
for direct $CP$ violation in $D\to K^+K^-$ and $D\to\pi^+\pi^-$
decays, with $CP$ asymmetries of ${\cal O}(10^{-2})$.
The current analysis also uses
$D\to K^+K^-$ and $D\to\pi^+\pi^-$ decays, and the above $CP$
asymmetries would bias our determinations of $\cos\delta$ and
$\sin\delta$ from these modes by ${\cal O}(10^{-2})$. By itself,
this bias is much smaller than our experimental uncertainties,
but its effect is further diluted by our use of additional $CP$
eigenstates in the analysis. $CP$ violation in
mixing and in the interference between mixing and decay would bias our
measured value of $y$. However, we are insensitive to these sources of
$CP$ violation at the levels currently allowed by experimental
constraints~\cite{hfag}.

The paper is organized as follows.  In Section~\ref{sec:formalism}, we
review the formalism of quantum-correlated $D^0\bar D^0$ decay.
Section~\ref{sec:selection} describes the event selection criteria and
$D$ reconstruction procedures.  The external measurements used
in the fit are summarized in Section~\ref{sec:externalMeas}.
Systematic uncertainties, which are also input to the fit, are discussed
in Section~\ref{sec:systematics}.  Finally, we present and discuss our main
fit results in Section~\ref{sec:results}.


\section{Formalism and Phenomenology}\label{sec:formalism}

For decays of isolated $D^0$ mesons, we define the following quantities for
each final state $i$:
\begin{eqnarray}
\label{eq:r2General}
  r_i^2 &\equiv& \frac{\int \bar A_i(\mathbf x) \bar A_i^*(\mathbf x) d\mathbf x}{\int A_i(\mathbf x) A_i^*(\mathbf x) d\mathbf x} \\
\label{eq:coherenceGeneral}
  R_i e^{-i\delta_i} &\equiv& \frac{\int \bar A_i(\mathbf x) A_i^*(\mathbf x) d\mathbf x}{ r_i \int A_i(\mathbf x) A_i^*(\mathbf x) d\mathbf x }
\end{eqnarray}
where
$A_i\equiv\langle i|D^0\rangle$ and $\bar A_i\equiv\langle i|\bar D^0\rangle$
are the amplitudes for the transitions of $D^0$ and $\bar D^0$, respectively,
to the final state $i$.
The integrals are taken over the phase space for mode $i$.
Thus, $\delta_i$ is an average phase for the final state
$i$, and $R_i\in [0,1]$ is a coherence factor~\cite{Atwood:2003mj}
that characterizes the variation of $\delta_i$ over phase space.
If the final state is two-body, like $K^-\pi^+$, then $\delta_i$ is constant
over phase space, and $R_i=1$.

For a $D^0\bar D^0$ pair produced through the $\psi(3770)$ resonance,
the decay rate to an exclusive final state $\{i,j\}$, where $i$ and $j$
label the final states of the two $D$ mesons, follows from the
antisymmetric amplitude ${\cal M}_{ij}$:
\begin{eqnarray}\label{eq:antisym}
\nonumber
\Gamma(i,j) \propto {\cal M}^2_{ij} &=&
        \left|A_i \bar A_j - \bar A_i A_j \right|^2 \\
\label{eq:rates}
&=& \left|\langle i|D_2\rangle\langle j|D_1\rangle -
        \langle i|D_1\rangle\langle j|D_2\rangle \right|^2 + {\cal O}(x^2, y^2),
\end{eqnarray}
where the ${\cal O}(x^2, y^2)$ term represents a mixed amplitude. The
interference
between mixed and unmixed amplitudes vanishes when integrated over time because
it depends on the difference between the $D^0$ and $\bar D^0$ decay times.
If we denote the charge conjugates of modes $i$ and $j$ by $\bar\imath$ and
$\bar\jmath$, then
Equation~(\ref{eq:antisym}) leads to the following expressions in terms of the
parameters defined above:
\begin{eqnarray}\label{eq:ratesCF}
\Gamma(i,\bar\jmath) = \Gamma(\bar\imath, j) &\propto& A_i^2 A_j^2 \left(
  1 + r_i^2 r_j^2 - 2 r_i R_i\cos\delta_i \ r_j R_j\cos\delta_j - 2 r_i R_i\sin\delta_i \ r_j R_j\sin\delta_j \right) \\
\label{eq:ratesDCS}
\Gamma(i,j) = \Gamma(\bar\imath, \bar\jmath) &\propto& A_i^2 A_j^2 \left(
  r_i^2 + r_j^2 - 2 r_i R_i\cos\delta_i \ r_j R_j\cos\delta_j + 2 r_i R_i\sin\delta_i \ r_j R_j\sin\delta_j \right),
\end{eqnarray}
where the latter rate is reduced by half if $i$ and $j$ are identical.
Experimentally, these rates correspond to yields of double tags (DT), which are
events where both $D^0$ and $\bar D^0$ are reconstructed.

The above amplitudes are normalized to the uncorrelated branching fractions
${\cal B}_i$:
\begin{eqnarray}
\label{eq:bfCF}
{\cal B}_i &\equiv& {\cal B}(D^0\to i) = A_i^2\left[ 1 + r_i R_i (y\cos\delta_i + x\sin\delta_i) \right] \\
\label{eq:bfDCS}
{\cal B}_{\bar\imath} &\equiv& {\cal B}(\bar D^0\to i) = A_i^2\left[ r_i^2 + r_i R_i (y\cos\delta_i - x\sin\delta_i) \right].
\end{eqnarray}
These ${\cal B}_i$ are related to rates of single tags (ST), or individually
reconstructed $D^0$ or $\bar D^0$ candidates, which are obtained by summing
over DT rates:
\begin{equation}\label{eq:ratesST}
\Gamma(i,X) = \sum_j\left[ \Gamma(i,j) + \Gamma(i,\bar\jmath) \right ] =
{\cal B}_i + {\cal B}_{\bar\imath} =
A_i^2 \left( 1 + 2 y r_i R_i \cos\delta_i + r_i^2 \right).
\end{equation}
Here, we have used an expression for $y$ in terms of $r_i$, $R_i$, and
$\delta_i$, which is derived from Eqs.~(\ref{eq:y}--\ref{eq:d2}) and
Eqs.~(\ref{eq:r2General}--\ref{eq:coherenceGeneral}):
\begin{equation}
y = \frac{ \sum_i \left[ \int | A_i(\mathbf x) - \bar A_i(\mathbf x) |^2 d\mathbf x -
\int | A_i(\mathbf x) + \bar A_i(\mathbf x) |^2 d\mathbf x \right]}
{\sum_i \left[ \int | A_i(\mathbf x) - \bar A_i(\mathbf x) |^2 d\mathbf x +
\int | A_i(\mathbf x) + \bar A_i(\mathbf x) |^2 d\mathbf x \right]}
= -2\frac{\sum_i A_i^2 r_i R_i \cos\delta_i}{\sum_i A_i^2 ( 1 + r_i^2 ) }.
\end{equation}
Thus, both ST rates and the total rate, $\Gamma_{D^0\bar D^0}$, are
unaffected by quantum correlations between the $D^0$ and $\bar D^0$ decays,
and our sensitivity to mixing comes from comparing ST to DT rates.

Table~\ref{tab:rdelta} gives the notation for the various $r_i$ and $\delta_i$
that appear in this analysis. The final states of mixed $CP$ (denoted by
$f$ and $\bar f$ below) that we consider are $K^\mp\pi^\pm$ and
$K^0_S\pi^+\pi^-$. Following Ref.~\cite{kspipi}, the $K^0_S\pi^+\pi^-$
Dalitz plot is divided into eight bins according to the strong phase of
the decay amplitude. We denote the portions of $K^0_S\pi^+\pi^-$ in
phase bin $i$ by $Y_i$ and $\bar Y_i$, where $m_{K^0_S\pi^-} < m_{K^0_S\pi^+}$
for $Y_i$, and $m_{K^0_S\pi^-} > m_{K^0_S\pi^+}$ for $\bar Y_i$.
The corresponding amplitude ratio magnitudes
and branching fraction ratios integrated over bin $i$ are denoted by
$\rho_i$ and $Q_i$, respectively.
As in Ref.~\cite{kspipi}, we denote the real and imaginary parts of
Eq.~(\ref{eq:coherenceGeneral}) by $c_i$ and $s_i$, respectively, but with the 
opposite sign convention for $s_i$.
Semileptonic final states ($\ell^\pm$),
$CP$-even eigenstates ($S_+$), and $CP$-odd eigenstates ($S_-$)
have known values of $r_i$ and $\delta_i$, which give them
unique leverage in determining the parameters in the other final states,
as demonstrated below. Note that, as shown in Eqs.~(\ref{eq:bfCF})
and~(\ref{eq:bfDCS}), the ratio ${\cal B}_{\bar\imath}/{\cal B}_i$ does not
equal $r_i^2$ in general.

\begin{table}[htb]
\begin{center}
\caption{Parameters describing the ratio of amplitudes $A_i$ and $\bar A_i$ for
the final states $i$.
The $\cdot$ indicates that we do not make explicit reference to
$\delta_i$ for the $Y_k$ modes in this article, but consider only
$c_k$ and $s_k$ instead.}
\label{tab:rdelta}
\begin{tabular}{cccccc}
\hline\hline
~Final State~ & ~$r_i$~ & ~$\delta_i$~ &
~$R_i \cos\delta_i$~ & ~$R_i\sin\delta_i$~ &
~${\cal B}_{\bar\imath}/{\cal B}_i$~ \\
\hline
$K^\mp\pi^\pm$ & $r$ & $\delta$ & $\cos\delta$ & $\sin\delta$ & $R_{\rm WS}$ \\
$Y_k/\bar Y_k$ & $\rho_k$ & $\cdot$ & $c_k$ & $s_k$ & $Q_k$ \\
$S_+$ & $1$ & $\pi$ & $-1$ & 0 & 1 \\
$S_-$ & $1$ & 0 & $+1$ & 0 & 1 \\
$\ell^\pm$ & 0 & --- & --- & --- & 0 \\
\hline\hline
\end{tabular}
\end{center}
\end{table}

Using the definitions in Table~\ref{tab:rdelta}, we evaluate
in Table~\ref{tab:rates} the quantum-correlated $D^0\bar D^0$
branching fractions, ${\cal F}^{\rm cor}$, for all categories
of final states reconstructed in this analysis;
we also give the corresponding uncorrelated branching fractions,
${\cal F}^{\rm unc}$.  Comparing ${\cal F}^{\rm cor}$ with ${\cal F}^{\rm unc}$
allows us to extract $y$, $r^2$, $\cos\delta$, and $\sin\delta$.
Although we neglect $x^2$ and $y^2$ terms in general, we report a result for
$x^2$ as determined solely from the suppressed
$\{K^\pm\pi^\mp, K^\pm\pi^\mp\}$ final states.

\begin{table}[htbp]
\begin{center}
\caption{Correlated ($C$-odd) and uncorrelated effective $D^0\bar D^0$
branching fractions, ${\cal F}^{\rm cor}$ and
${\cal F}^{\rm unc}$, to leading order in $x$,
$y$ and $R_{\rm WS}$, divided by ${\cal B}_i$
for ST modes $i$ and
${\cal B}_i{\cal B}_j$ for DT modes $\{i, j\}$.
Charge conjugate modes are implied.}
\label{tab:rates}
\begin{tabular}{ccc}
\hline\hline
Mode & Correlated & Uncorrelated \\
\hline
$K^-\pi^+$ &
        $1+R_{\rm WS}$ &
        $1+R_{\rm WS}$ \\
$S_+$ & $2$ & $2$ \\
$S_-$ & $2$ & $2$ \\
$Y_k$ & $1+Q_k$ & $1+Q_k$ \\
$K^-\pi^+$, $K^-\pi^+$ &
        $R_{\rm M} [(1+R_{\rm WS})^2-4r\cos\delta(r\cos\delta+y)]$ &
        $R_{\rm WS}$ \\
$K^-\pi^+$, $K^+\pi^-$ &
$(1+R_{\rm WS})^2-4r\cos\delta(r\cos\delta+y)$ &
        $1+R_{\rm WS}^2$ \\
$K^-\pi^+$, $S_+$ &
        $1+ R_{\rm WS}+2r\cos\delta+y$ &
        $1+R_{\rm WS}$ \\
$K^-\pi^+$, $S_-$ &
        $1+R_{\rm WS}- 2r\cos\delta-y$ &
        $1+R_{\rm WS}$ \\
$K^-\pi^+$, $\ell^-$ &
        $1-ry\cos\delta-rx\sin\delta$ &
        $1$ \\
$K^-\pi^+$, $\ell^+$ &
        $r^2(1-ry\cos\delta-rx\sin\delta)$ &
        $R_{\rm WS}$ \\
$K^-\pi^+$, $\bar Y_{i}$ &
        $\begin{array}{c}(1+R_{\rm WS})(1+Q_i)-r^2-\rho_i^2 \\ -2(r\cos\delta+y)(\rho_ic_i+y) +2r\sin\delta \rho_i s_i\end{array}$ &
        $1 + R_{\rm WS}Q_i$ \\
$K^-\pi^+$, $Y_{i}$ &
        $\begin{array}{c}(1+R_{\rm WS})(1+Q_i)-1-r^2 \rho_i^2 \\ -2(r\cos\delta+y)(\rho_i c_i+y) -2r\sin\delta \rho_i s_i\end{array}$ &
        $R_{\rm WS}+Q_i$ \\
$S_+$, $S_+$ & 0 & $1$ \\
$S_-$, $S_-$ & 0 & $1$ \\
$S_+$, $S_-$ &
        $4$ &
        $2$ \\
$S_+$, $\ell^-$ &
        $1+y$ &
        $1$ \\
$S_-$, $\ell^-$ &
        $1-y$ &
        $1$\\
$S_+$, $Y_i$ & $1+ Q_i+2 \rho_i c_i +y$ &
        $1+Q_i$ \\
$S_-$, $Y_i$ & $1+ Q_i-2 \rho_i c_i -y$ &
        $1+Q_i$ \\
$Y_i$, $\ell^-$ &
        $1- \rho_i y c_i - \rho_i x s_i$ &
        $1$ \\
$Y_i$, $\ell^+$ &
        $\rho_i^2( 1- \rho_i y c_i - \rho_i x s_i )$ &
        $Q_i$ \\
$Y_{i}$, $\bar Y_{j}$ &
        $\begin{array}{c}(1+Q_i)(1+Q_j)-\rho_i^2-\rho_j^2 \\ -2(\rho_i c_i+y)(\rho_j c_j+y) +2\rho_is_i \rho_js_j\end{array}$ &
        $1 + Q_iQ_j$ \\
$Y_{i}$, $Y_{j}$ &
        $\begin{array}{c}(1+Q_i)(1+Q_j)-1-\rho_i^2\rho_j^2 \\ -2(\rho_ic_i+y)(\rho_jc_j+y) -2\rho_is_i \rho_js_j\end{array}$ &
        $Q_i + Q_j$ \\
\hline\hline
\end{tabular}
\end{center}
\end{table}

From Table~\ref{tab:rates}, one finds that, given $r^2$ and $y$,
$\cos\delta$ can be determined by measuring the size of the
interference between $K^-\pi^+$ and a $CP$ eigenstate.
The $CP$ of the eigenstate tags the $K^-\pi^+$
parent $D$ to be a $CP$ eigenstate with the opposite eigenvalue.
Since this $D$ eigenstate is a linear combination of the
flavor eigenstates $D^0$ and $\bar D^0$, the decay rate is
modulated by the relative phase between the $D^0\to K^-\pi^+$
and $\bar D^0\to K^-\pi^+$ amplitudes.

Similarly, probing $\sin\delta$ requires the
interference of $K^-\pi^+$ with another mode, such as $K^0_S\pi^+\pi^-$,
that has non-zero $R_i\sin\delta_i$.
However, unlike $CP$ eigenstates, the phases in $K^0_S\pi^+\pi^-$ are not
fixed by a fundamental symmetry, so we must measure
$\sin\delta$ and $s_i$ simultaneously.
Since these sine factors only appear in
products with other sine factors, there is an
overall sign ambiguity, which can be resolved by combining our measurements
of $\sin\delta$ and $\cos\delta$ with external measurements of $y'$ and $x$.

Our main source of information on $y$ comes from $CP$-tagged semileptonic
decays.  In these weak transitions, the semileptonic decay width is independent
of the parent $D$ meson's $CP$ eigenvalue. In contrast, the total
width of the parent meson reflects its $CP$ eigenvalue:
$\Gamma_{1,2}=\Gamma(1 \mp y)$,
so the semileptonic branching fraction for $D_1$ or $D_2$ is modified by a
factor of $1\pm y$.
Thus, we determine $y$ using exclusive final states $\{S_\pm, \ell\}$,
where the $S_\pm$
identifies the $CP$ eigenvalue of the semileptonic decay's parent $D$.
In this case, summing $\ell^+$ and $\ell^-$ rates gives
${\cal F}^{\rm cor}_{S_\pm, \ell} \approx 2{\cal B}_{S_\pm}{\cal B}_{\ell}(1\pm y)$.

For $r^2$, we use the fact that, because of the vanishing interference
between mixed and unmixed amplitudes, a DT with a semileptonic $K\ell\nu_\ell$
decay probes the bare matrix element squared, not the
branching fraction, of the partner $D$. Therefore, we determine
$r^2$ directly from $\{K\pi, K\ell\nu_\ell\}$ DT modes by taking the yield ratio
for combinations with same-sign kaons and opposite-sign kaons.


\section{Event Selection and Reconstruction}~\label{sec:selection}

Our current analysis is based on the full CLEO-c $\psi(3770)$ dataset
with an integrated luminosity of 818 ${\rm pb}^{-1}$,
collected with the CLEO-c
detector, which is described in Refs.~\cite{cleo2,cleo2.5,cleo3,rich,cleo-c}.
We estimate signal efficiencies, background contributions, and probabilities
for misreconstructing a produced signal decay in a different signal mode
(crossfeed) using a GEANT-based~\cite{geant} Monte Carlo simulated
sample of uncorrelated $D^0\bar D^0$ decays with an effective integrated
luminosity 20 times larger than that of our data sample.
We reconstruct the final states shown in Table~\ref{tab:finalStates},
with $\pi^0\to\gamma\gamma$, $\eta\to\gamma\gamma$, $K^0_S\to\pi^+\pi^-$, and
$\omega\to\pi^+\pi^-\pi^0$. Final states without $K^0_L$ mesons and
neutrinos are fully reconstructed.
For modes with $K^0_L$ mesons and neutrinos, which generally do not
interact with the detector, we use a partial reconstruction technique,
inferring the presence of the undetected particle via conservation of
energy and momentum.
In specifying the $CP$ eigenvalue of a final state,
we neglect $CP$ violation in $K^0$ decays.

\begin{table}[htb]
\caption{$D$ final states reconstructed in this analysis.}
\label{tab:finalStates}
\begin{tabular}{ccc}
\hline\hline
Type & Reconstruction & Final States \\
\hline
$f$ & full &
$K^-\pi^+$, $Y_0-Y_7$ \\
$\bar f$ & full &
$K^+\pi^-$, $\bar Y_0-\bar Y_7$ \\
$S_+$ & full &
$K^+K^-$, $\pi^+\pi^-$, $K^0_S\pi^0\pi^0$ \\
$S_+$ & partial &
$K^0_L\pi^0$, $K^0_L\eta$, $K^0_L\omega$ \\
$S_-$ & full &
$K^0_S\pi^0$, $K^0_S\eta$, $K^0_S\omega$ \\
$S_-$ & partial &
$K^0_L\pi^0\pi^0$ \\
$\ell^+$ & partial &
$K^- e^+\nu_e$, $K^-\mu^+\nu_\mu$ \\
$\ell^-$ & partial &
$K^+ e^-\bar\nu_e$, $K^+\mu^-\bar\nu_\mu$ \\
\hline\hline
\end{tabular}
\end{table}

\begin{figure}[htbp]
\includegraphics*[width=\linewidth]{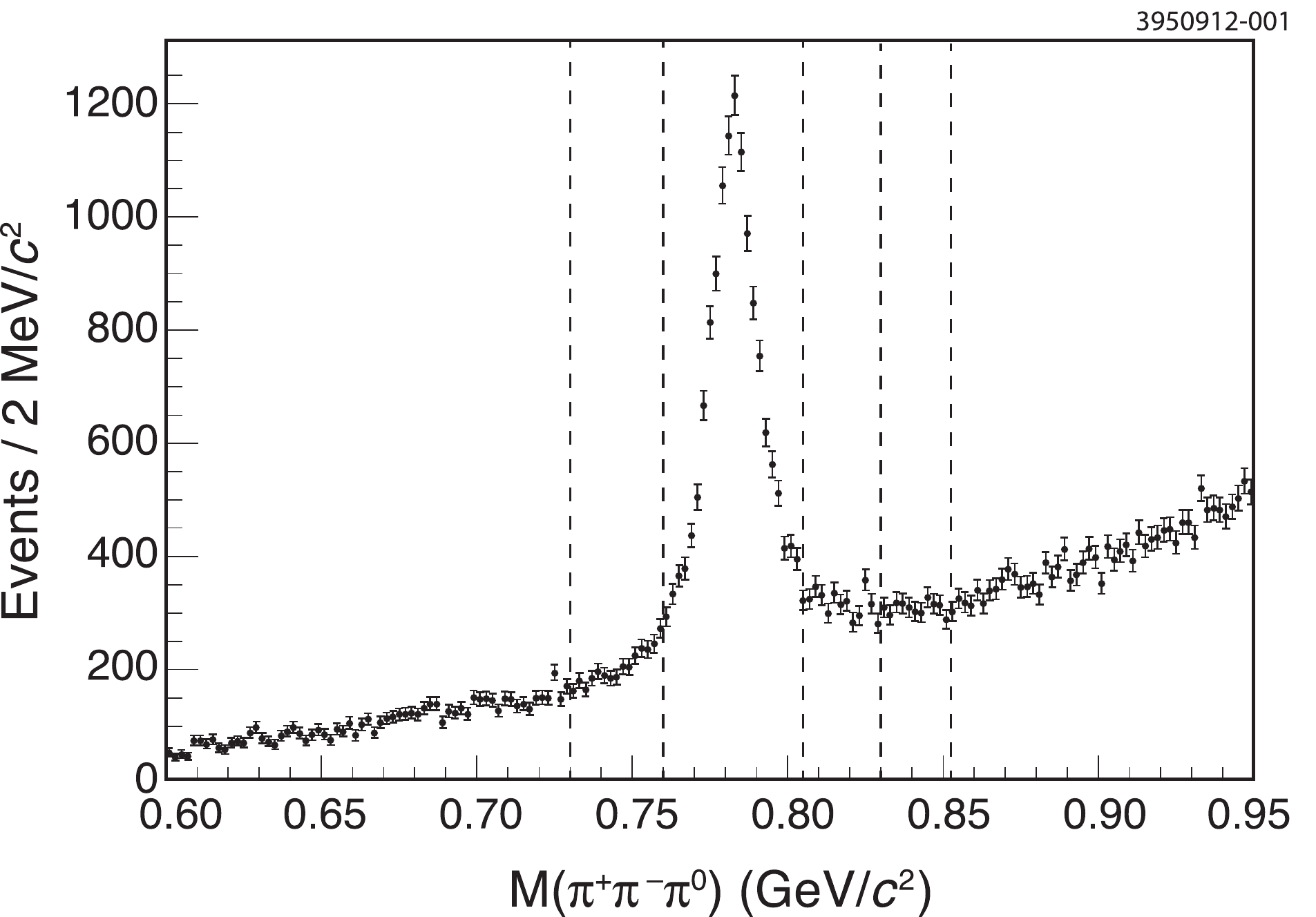}
\caption{Distribution of $M(\pi^+\pi^-\pi^0)$ for $\omega$ candidates.
Data are shown as points with error bars, and the dashed
lines mark the boundaries of the signal and sideband regions.
As indicated in the text, the lower sideband ends at the lower limit
of the graph.
}
\label{fig:omega}
\end{figure}

Final states that are common to those used in Ref.~\cite{tqca1} are
reconstructed with the same methods and selection criteria,
except where noted below. In particular, the selection of $\pi^\pm$,
$K^\pm$, and $K^0_S$ candidates remains unchanged. For $\pi^0$ and $\eta$
candidates, we loosen the shower shape requirements to improve
the agreement between efficiencies in data and those in simulated events.
For all modes with $\omega$ candidates, we now apply a
sideband subtraction in the $M(\pi^+\pi^-\pi^0)$ spectrum.
Figure~\ref{fig:omega} shows the invariant mass distribution of
$\omega$ candidates, along with the signal region of
$760.0 \ {\rm MeV}/c^2 < M(\pi^+\pi^-\pi^0) < 805.0 \ {\rm MeV}/c^2$ and
sideband regions of
$600.0 \ {\rm MeV}/c^2 < M(\pi^+\pi^-\pi^0) < 730.0 \ {\rm MeV}/c^2$
and $830.0 \ {\rm MeV}/c^2 < M(\pi^+\pi^-\pi^0) < 852.5 \ {\rm MeV}/c^2$.
The limited range of the upper sideband is chosen to minimize the
effect of $\rho^0\to\pi^+\pi^-$ and $\rho^\pm\to\pi^\pm\pi^0$ decays,
which alter the shape of the background for
$M(\pi^+\pi^-\pi^0)$ greater than approximately $870 \ {\rm MeV}/c^2$.
The sidebands are scaled by a factor determined by fitting the
$M(\pi^+\pi^-\pi^0)$ distribution in simulated events to a signal
Gaussian plus a polynomial background and integrating the fitted
background function.
This sideband subtraction eliminates peaking backgrounds, which
accounted for $5-10\%$ of the observed $\omega$ yields in
Ref.~\cite{tqca1}.
We also make use of $K^0_S\pi^+\pi^-$ DT yields, efficiencies, and
background estimates
from Ref.~\cite{kspipi}, for the subset of modes in that analysis without
$K^0_L\pi^+\pi^-$, $K^-\pi^+\pi^0$, or $K^-\pi^+\pi^-\pi^+$.

\subsection{Single Tags}\label{sec:st}

\begin{figure}[htb]
\includegraphics*[width=\linewidth]{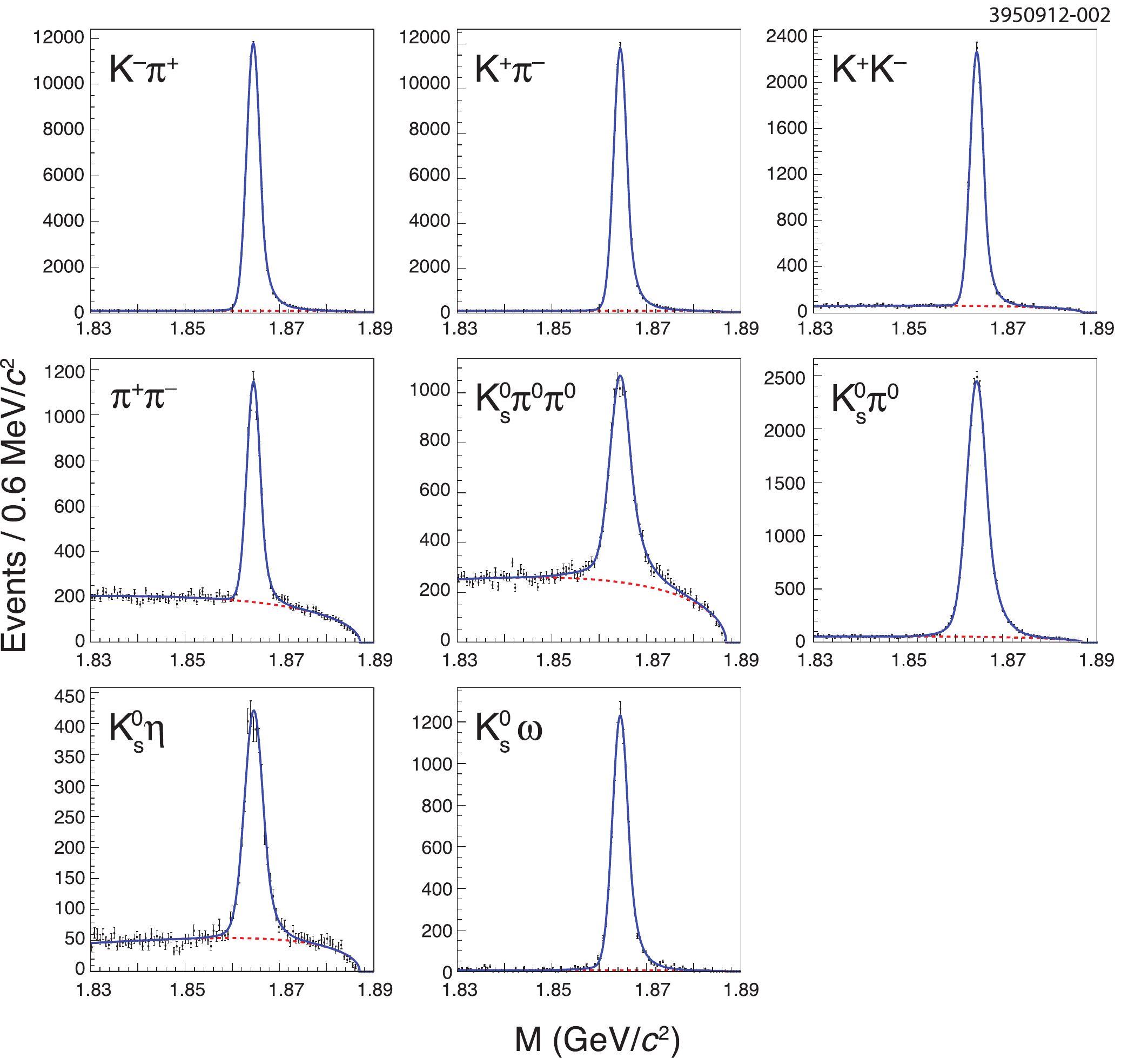}
\caption{
ST $M$ distributions and fits.
Data are shown as points with error bars. The solid lines show
the total fits, and the dashed lines show the background components.
}
\label{fig:st}
\end{figure}

We reconstruct ST candidates for the 8 modes in Table~\ref{tab:DeltaEcuts},
utilizing the technique described in Ref.~\cite{tqca1}.
We do not include ST yields for $K^0_S\pi^+\pi^-$.
As before, we identify ST candidates using two kinematic variables:
the beam-constrained candidate mass $M$ and the energy difference
$\Delta E$, which are defined to be
\begin{eqnarray}
M &\equiv& \sqrt{ E_0^2 / c^4 - {\mathbf p}_D^2 / c^2 } \\
\Delta E &\equiv& E_D - E_0,
\end{eqnarray}
where ${\mathbf p}_D$ and $E_D$ are the total momentum and energy of the $D$
candidate, and $E_0$ is the beam energy.
After applying the mode-dependent requirements on $\Delta E$ listed in
Table~\ref{tab:DeltaEcuts}, we determine the ST yields by fitting the $M$
distributions, shown in Fig.~\ref{fig:st}, to a signal shape derived from
simulated signal events and to
a background ARGUS function~\cite{argus}.

\begin{table}[htb]
\caption{Requirements on $\Delta E$ for ST $D$ candidates.
\label{tab:DeltaEcuts}}
\begin{tabular}{lc}\hline\hline
Mode              &  Requirement (GeV)\\ \hline
$K^-\pi^+$        &  $|\Delta E|<0.0294$      \\
$K^+\pi^-$        &  $|\Delta E|<0.0294$      \\
$K^+K^-$          &  $|\Delta E|<0.0200$    \\
$\pi^+\pi^-$      &  $|\Delta E|<0.0300$    \\
$K^0_S\pi^0\pi^0$ &  $-0.0550<\Delta E<0.0450$    \\
$K^0_S\pi^0$      &  $-0.0710<\Delta E<0.0450$    \\
$K^0_S\eta$       &  $-0.0550<\Delta E<0.0350$    \\
$K^0_S\omega$     &  $|\Delta E|<0.0250$    \\
\hline\hline
\end{tabular}
\end{table}

The measured ST yields and efficiencies are given in
Table~\ref{tab:STYieldEffs}. All efficiencies in this article include
constituent branching fractions. The yield
uncertainties are statistical and uncorrelated systematic, respectively.
The latter arise from modeling of multiple candidates in simulation and
variations in the signal lineshape.  Correlated systematic uncertainties
are discussed separately in Section~\ref{sec:systematics}.

\begin{table}[htbp]
\begin{center}
\caption{ST yields and efficiencies including constituent branching fractions.
Yield
uncertainties are statistical and uncorrelated systematic, respectively,
and efficiency uncertainties are statistical only.}
\label{tab:STYieldEffs}
\begin{tabular}{lcc}
\hline\hline
Mode &  ~~Yield~~ & ~~Efficiency (\%) \\ \hline
$K^-\pi^+$ &
   $75472\pm 300\pm 26$ &
   $63.74\pm 0.03$ \\
$K^+\pi^-$ &
   $75655\pm 299\pm 26$ &
   $64.76\pm 0.03$ \\
$K^+K^-$ &
   $13813\pm 134\pm 5$ &
   $56.15\pm 0.07$ \\
$\pi^+\pi^-$ &
   $6158\pm 114\pm 9$ &
   $72.08\pm 0.11$ \\
$K^0_S\pi^0\pi^0$ &
   $9209\pm 172\pm 16$ &
   $14.34\pm 0.04$ \\
$K^0_S\pi^0$ &
   $23025\pm 174\pm 17$ &
   $31.53\pm 0.04$ \\
$K^0_S\eta$ &
   $3251\pm 81\pm 17$ &
   $10.81\pm 0.05$ \\
$K^0_S\omega$ &
   $9292\pm 105\pm 7$ &
   $12.89\pm 0.03$ \\
\hline\hline
\end{tabular}
\end{center}
\end{table}

\subsection{Fully Reconstructed Hadronic Double Tags}\label{sec:dt}

\begin{figure}[htb]
\includegraphics*[width=\linewidth]{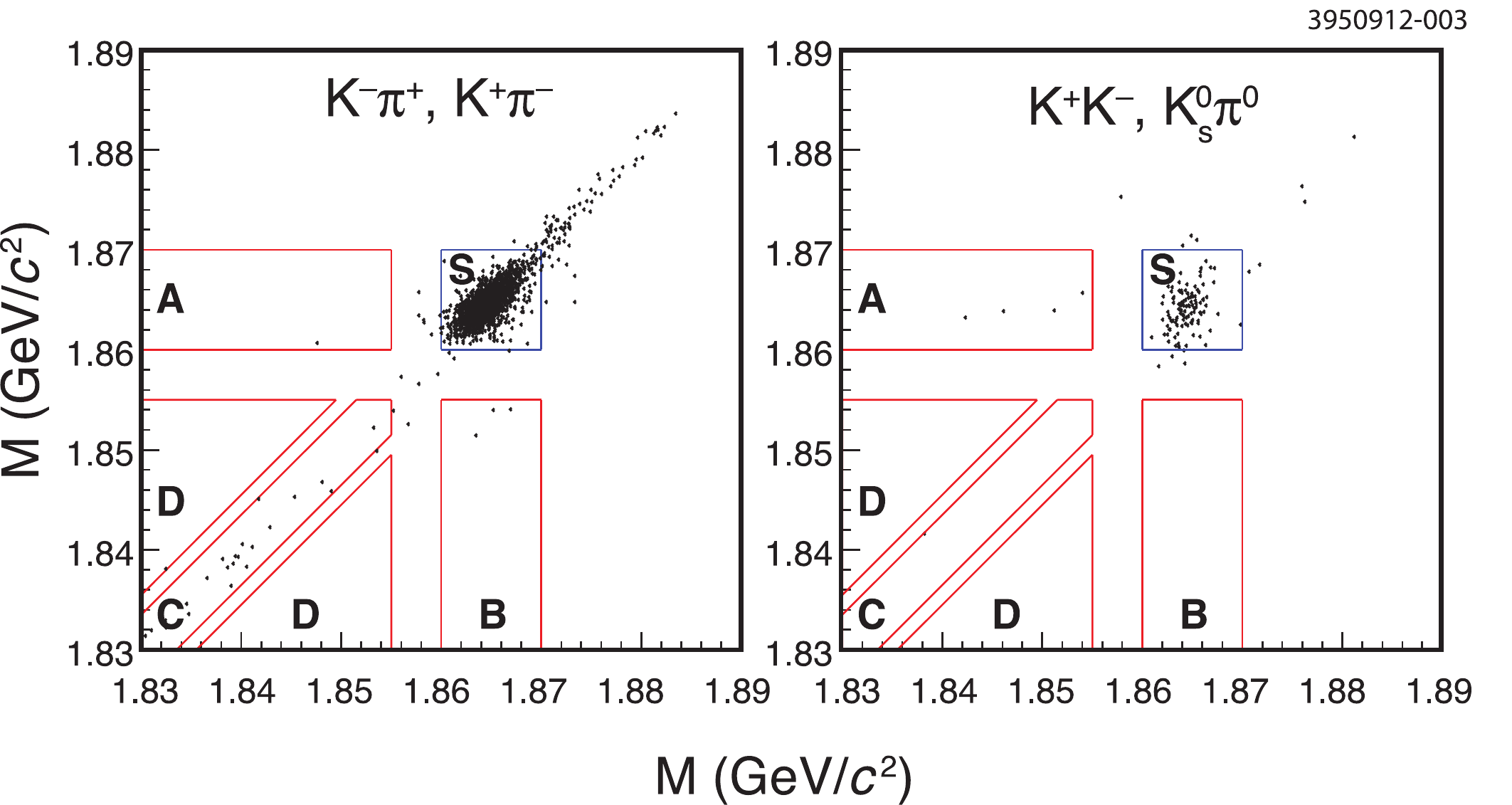}
\caption{Two-dimensional $M$ distributions with signal (S) and
sideband (A, B, C, D) regions depicted, for $\{K^-\pi^+, K^+\pi^-\}$
and $\{K^+K^-, K^0_S\pi^0\}$.
}
\label{fig:dt}
\end{figure}

We reconstuct two categories of DT final states: 24 $CP$-allowed combinations
of the 8 ST modes, and 136 modes with $K^0_S\pi^+\pi^-$. The second category
includes 64 measurements where $Y_i$ and $\bar Y_i$ are paired with
each of the eight ST modes [the sum of $\{K^-\pi^+, \bar Y_i\}$ and
$\{K^+\pi^-, Y_i\}$ (Cabibbo-favored), the sum of $\{K^-\pi^+, Y_i\}$ and
$\{K^+\pi^-, \bar Y_i\}$ (Cabibbo-suppressed), and the sum of $Y_i$ and
$\bar Y_i$ paired with the six fully-reconstructed $CP$ eigenstates], as well
as 36 Cabibbo-favored combinations
$\{Y_i, \bar Y_j\}+\{\bar Y_i, Y_j\}$, and 36
Cabibbo-suppressed combinations $\{Y_i, Y_j\}+\{\bar Y_i, \bar Y_j\}$.
The 24 DT modes in the first category above were also used in
Ref.~\cite{tqca1},
and we apply the same candidate selection criteria and yield determination
methods as before, with the addition of the $\omega$ mass sideband
subtraction discussed above.
Figure~\ref{fig:dt} shows some representative two-dimensional $M$
distributions. Event counts in the signal regions (S) are corrected by
background estimates from the sideband regions (A, B, C, D), with sideband
scaling factors determined by fitting the ST distributions in simulated
events in the same manner as Section~\ref{sec:st} and integrating the
fitted background function.
Table~\ref{tab:DTYieldEffs} gives the fully reconstructed DT yields and
efficiencies for modes without $K^0_S\pi^+\pi^-$.
To account for
systematic effects in the sideband definitions and in the extrapolation to the
signal regions, we assign a 100\% systematic uncertainty on the size of the
sideband subtractions, which is much smaller than the statistical
uncertainties in all cases.

For DT modes with $K^0_S\pi^+\pi^-$, we use the signal yields, efficiencies,
and background estimates determined in Ref.~\cite{kspipi}.

\begin{table}[htbp]
\begin{center}
\caption{Fully reconstructed DT yields and efficiencies
including constituent branching fractions,
for modes without $K^0_S\pi^+\pi^-$.
Yield uncertainties are statistical and uncorrelated systematic, respectively,
and efficiency uncertainties are statistical only.}
\label{tab:DTYieldEffs}
\begin{tabular}{lcc}
\hline\hline
Mode & Yield & Efficiency (\%) \\
\hline
$K^-\pi^+, K^-\pi^+$ &
   $5.6\pm 2.5\pm 0.4$ &
   $41.5\pm 2.8$ \\
$K^-\pi^+, K^+\pi^-$ &
   $1731\pm 42\pm 11$ &
   $40.0\pm 0.2$ \\
$K^-\pi^+, K^+K^-$ &
   $202\pm 14\pm 4$ &
   $35.2\pm 0.5$ \\
$K^-\pi^+, \pi^+\pi^-$ &
   $82.6\pm 9.1\pm 0.4$ &
   $44.5\pm 0.9$ \\
$K^-\pi^+, K^0_S\pi^0\pi^0$ &
   $132\pm 12\pm 1$ &
   $8.6\pm 0.2$ \\
$K^-\pi^+, K^0_S\pi^0$ &
   $252\pm 16\pm 1$ &
   $19.4\pm 0.3$ \\
$K^-\pi^+, K^0_S\eta$ &
   $36.7\pm 6.2\pm 1.3$ &
   $6.9\pm 0.3$ \\
$K^-\pi^+, K^0_S\omega$ &
   $109\pm 11\pm 1$ &
   $8.5\pm 0.2$ \\
$K^+\pi^-, K^+\pi^-$ &
   $4.0\pm 2.0\pm 0.0$ &
   $42.9\pm 2.9$ \\
$K^+\pi^-, K^+K^-$ &
   $191\pm 14\pm 1$ &
   $35.3\pm 0.5$ \\
$K^+\pi^-, \pi^+\pi^-$ &
   $77.3\pm 8.9\pm 0.7$ &
   $45.6\pm 0.9$ \\
$K^+\pi^-, K^0_S\pi^0\pi^0$ &
   $121\pm 11\pm 2$ &
   $9.1\pm 0.2$ \\
$K^+\pi^-, K^0_S\pi^0$ &
   $242\pm 16\pm 0$ &
   $20.0\pm 0.3$ \\
$K^+\pi^-, K^0_S\eta$ &
   $35.2\pm 6.0\pm 0.8$ &
   $6.9\pm 0.3$ \\
$K^+\pi^-, K^0_S\omega$ &
   $89.4\pm 10.2\pm 1.3$ &
   $8.7\pm 0.2$ \\
$K^+K^-, K^0_S\pi^0$ &
   $107\pm 11\pm 2$ &
   $18.1\pm 0.5$ \\
$K^+K^-, K^0_S\eta$ &
   $24.6\pm 5.0\pm 0.4$ &
   $5.6\pm 0.6$ \\
$K^+K^-, K^0_S\omega$ &
   $47.6\pm 7.2\pm 0.0$ &
   $7.2\pm 0.4$ \\
$\pi^+\pi^-, K^0_S\pi^0$ &
   $37.0\pm 6.1\pm 0.0$ &
   $21.3\pm 0.9$ \\
$\pi^+\pi^-, K^0_S\eta$ &
   $6.0\pm 2.5\pm 0.0$ &
   $6.6\pm 1.0$ \\
$\pi^+\pi^-, K^0_S\omega$ &
   $19.0\pm 4.7\pm 0.0$ &
   $9.4\pm 0.7$ \\
$K^0_S\pi^0\pi^0, K^0_S\pi^0$ &
   $53.0\pm 7.3\pm 0.0$ &
   $4.1\pm 0.2$ \\
$K^0_S\pi^0\pi^0, K^0_S\eta$ &
   $10.0\pm 3.2\pm 0.0$ &
   $1.4\pm 0.2$ \\
$K^0_S\pi^0\pi^0, K^0_S\omega$ &
   $18.0\pm 4.8\pm 1.0$ &
   $1.5\pm 0.1$ \\
\hline\hline
\end{tabular}
\end{center}
\end{table}

\subsection{\boldmath Double Tags with $K^0_L$}

For hadronic DT modes with a single $K^0_L$, we employ the same partial
reconstruction technique as for $K^0_L\pi^0$ decays in Ref.~\cite{tqca1},
where the $K^0_L$ is identified by the four-vector
recoiling against all other observed particles in the event.
In the current analysis, we tag $K^0_L\pi^0$, $K^0_L\eta$,
$K^0_L\omega$, and $K^0_L\pi^0\pi^0$ decays
with fully reconstructed ST candidates, as allowed by $CP$ conservation
and selected as described in
Section~\ref{sec:st}, and with an additional requirement of
$1.86 \ {\rm GeV}/c^2 < M < 1.87 \ {\rm GeV}/c^2$.
Each ST candidate is combined with a $\pi^0$, $\eta$, $\omega$
candidate, or a pair of $\pi^0$ candidates.
The signal process with $K^0_L$ appears as a peak in the
squared recoil mass, $M^2_{\rm miss}$, against this system.
As in Ref.~\cite{tqca1}, we suppress the background by vetoing events with
additional unassigned charged particles, but we veto addtional $\pi^0$
candidates only for the $K^0_L\eta$ mode. In addition, for all
$K^0_L$ modes, we follow Ref.~\cite{kspipi} by applying a veto on extra
showers outside an energy-dependent cone around the predicted $K^0_L$
direction.

\begin{figure}[htb]
\includegraphics*[width=\linewidth]{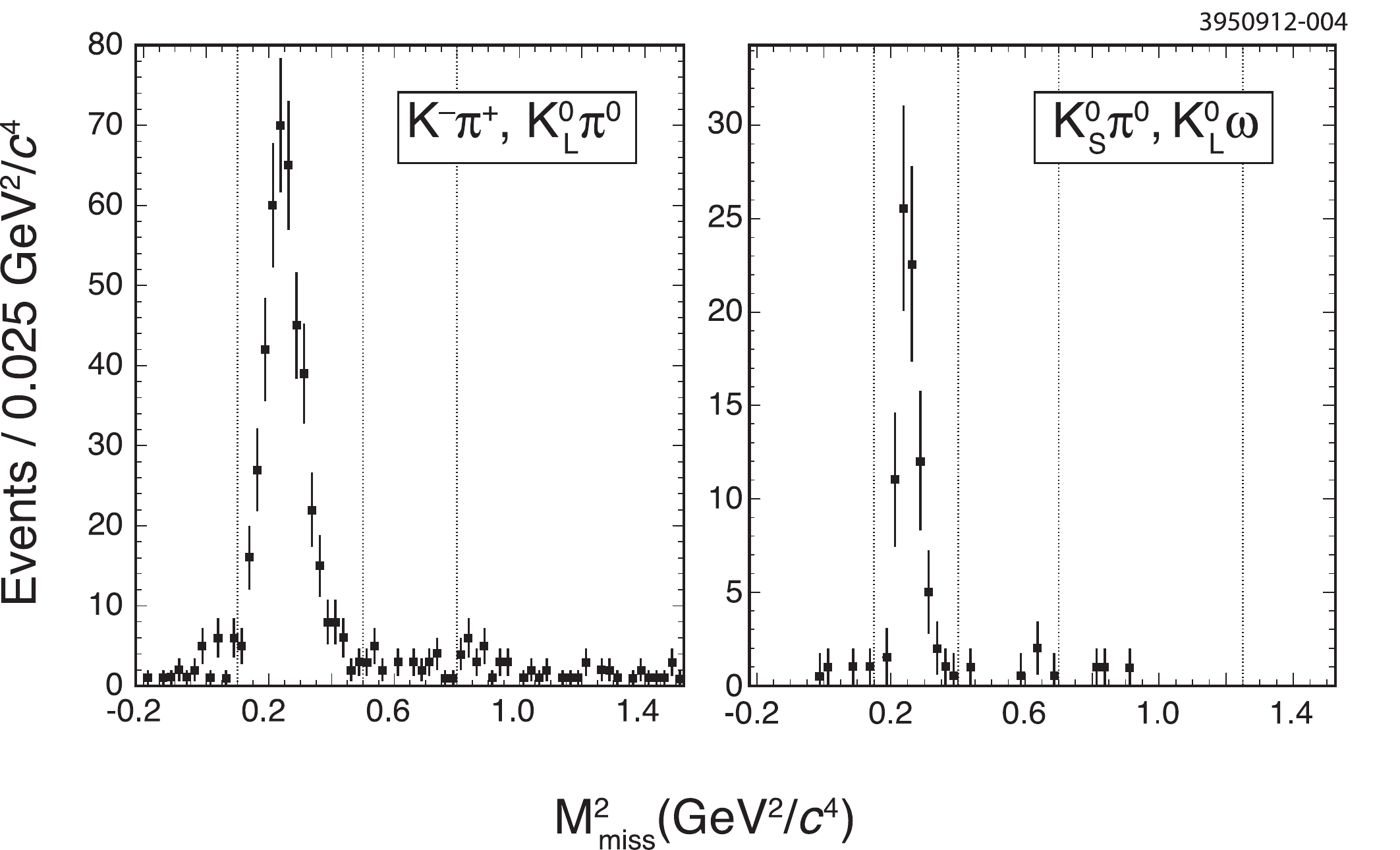}
\caption{
Distributions of $M^2_{\rm miss}$ for $\{K^-\pi^+, K^0_L\pi^0\}$
and $\{K^0_S\pi^0, K^0_L\omega\}$.
The dashed lines show the signal and sideband regions.
As indicated in Table~\ref{tab:klxRegions}, the sideband for
$\{K^-\pi^+, K^0_L\pi^0\}$ extends beyond the upper limit of the graph.
}
\label{fig:kl}
\end{figure}

Figure~\ref{fig:kl} shows examples of the resultant $M^2_{\rm miss}$
distributions in data. We obtain yields from event counts in the
signal and sideband regions as shown in Table~\ref{tab:klxRegions}, where
the sideband is scaled by a factor determined from simulated events.
We also subtract a small contribution due to continuum $q\bar q$ production,
which is also estimated from simulated events.

\begin{table}[htb]
\caption{Signal and sideband regions in $M^2_{\rm miss}$ (${\rm GeV}^2/c^4$)
for $K^0_L$ modes.}
\label{tab:klxRegions}
\begin{tabular}{lcc}\hline\hline
Mode              &  Signal Region & Sideband Region \\
\hline
$K^0_L\pi^0$ & [ 0.10, 0.50 ] & [ 0.80, 2.00 ] \\
$K^0_L\eta$ & [ 0.10, 0.45 ] & [ 0.75, 1.75 ] \\
$K^0_L\omega$ & [ 0.15, 0.40 ] & [ 0.70, 1.25 ] \\
$K^0_L\pi^0\pi^0$ & [ 0.10, 0.50 ] & [ 0.80, 1.80 ] \\
\hline\hline
\end{tabular}
\end{table}

Table~\ref{tab:K0LYieldEffs} lists the DT yields and efficiencies for
$K^0_L$ modes without $K^0_S\pi^+\pi^-$.
There are no uncorrelated systematic uncertainties for these modes.
In the fit, we also include the $\{K^0_S\pi^+\pi^-, K^0_L\pi^0\}$ measurements
from Ref.~\cite{kspipi}, which uses the same technique and criteria for
$K^0_L\pi^0$ as those described above.

\begin{table}[htbp]
\begin{center}
\caption{DT yields and efficiencies including constituent branching fractions,
for $K^0_L$ modes without
$K^0_S\pi^+\pi^-$. Uncertainties are statistical only. All systematic
uncertainties are correlated and are discussed in 
Section~\ref{sec:systematics}.}
\label{tab:K0LYieldEffs}
\begin{tabular}{lcc}
\hline\hline
Mode & Yield & Efficiency (\%) \\
\hline
$K^-\pi^+, K^0_L\pi^0$ &
   $425\pm 21$ &
   $29.9\pm 0.2$ \\
$K^+\pi^-, K^0_L\pi^0$ &
   $381\pm 20$ &
   $31.1\pm 0.2$ \\
$K^0_S\pi^0, K^0_L\pi^0$ &
   $235\pm 15$ &
   $14.6\pm 0.1$ \\
$K^0_S\eta, K^0_L\pi^0$ &
   $28.0\pm 5.4$ &
   $5.34\pm 0.08$ \\
$K^0_S\omega, K^0_L\pi^0$ &
   $60.8\pm 8.7$ &
   $5.33\pm 0.07$ \\
$K^-\pi^+, K^0_L\eta$ &
   $70.8\pm 8.6$ &
   $11.2\pm 0.1$ \\
$K^+\pi^-, K^0_L\eta$ &
   $53.7\pm 7.6$ &
   $11.5\pm 0.1$ \\
$K^0_S\pi^0, K^0_L\eta$ &
   $21.7\pm 4.8$ &
   $5.55\pm 0.08$ \\
$K^0_S\eta, K^0_L\eta$ &
   $7.6\pm 2.8$ &
   $2.01\pm 0.05$ \\
$K^0_S\omega, K^0_L\eta$ &
   $9.3\pm 3.5$ &
   $2.04\pm 0.05$ \\
$K^-\pi^+, K^0_L\omega$ &
   $143\pm 13$ &
   $12.1\pm 0.1$ \\
$K^+\pi^-, K^0_L\omega$ &
   $155\pm 14$ &
   $12.2\pm 0.1$ \\
$K^0_S\pi^0, K^0_L\omega$ &
   $80.7\pm 9.8$ &
   $5.70\pm 0.08$ \\
$K^0_S\eta, K^0_L\omega$ &
   $5.9\pm 3.2$ &
   $2.06\pm 0.05$ \\
$K^0_S\omega, K^0_L\omega$ &
   $27.5\pm 5.6$ &
   $1.86\pm 0.05$ \\
$K^-\pi^+, K^0_L\pi^0\pi^0$ &
   $157\pm 13$ &
   $13.0\pm 0.1$ \\
$K^+\pi^-, K^0_L\pi^0\pi^0$ &
   $133\pm 12$ &
   $13.2\pm 0.1$ \\
$K^+K^-, K^0_L\pi^0\pi^0$ &
   $57.1\pm 7.7$ &
   $10.9\pm 0.1$ \\
$\pi^+\pi^-, K^0_L\pi^0\pi^0$ &
   $14.3\pm 4.9$ &
   $14.5\pm 0.1$ \\
$K^0_S\pi^0\pi^0, K^0_L\pi^0\pi^0$ &
   $36.6\pm 6.5$ &
   $2.85\pm 0.06$ \\
\hline\hline
\end{tabular}
\end{center}
\end{table}

\subsection{Semileptonic Double Tags}

This Section describes our reconstruction of semileptonic $D$
decays paired with fully reconstructed hadronic tags. In
Section~\ref{sec:twomissingparticles} below, we discuss an additional
semileptonic mode with two undetected particles.

\begin{figure}[htb]
\includegraphics*[width=\linewidth]{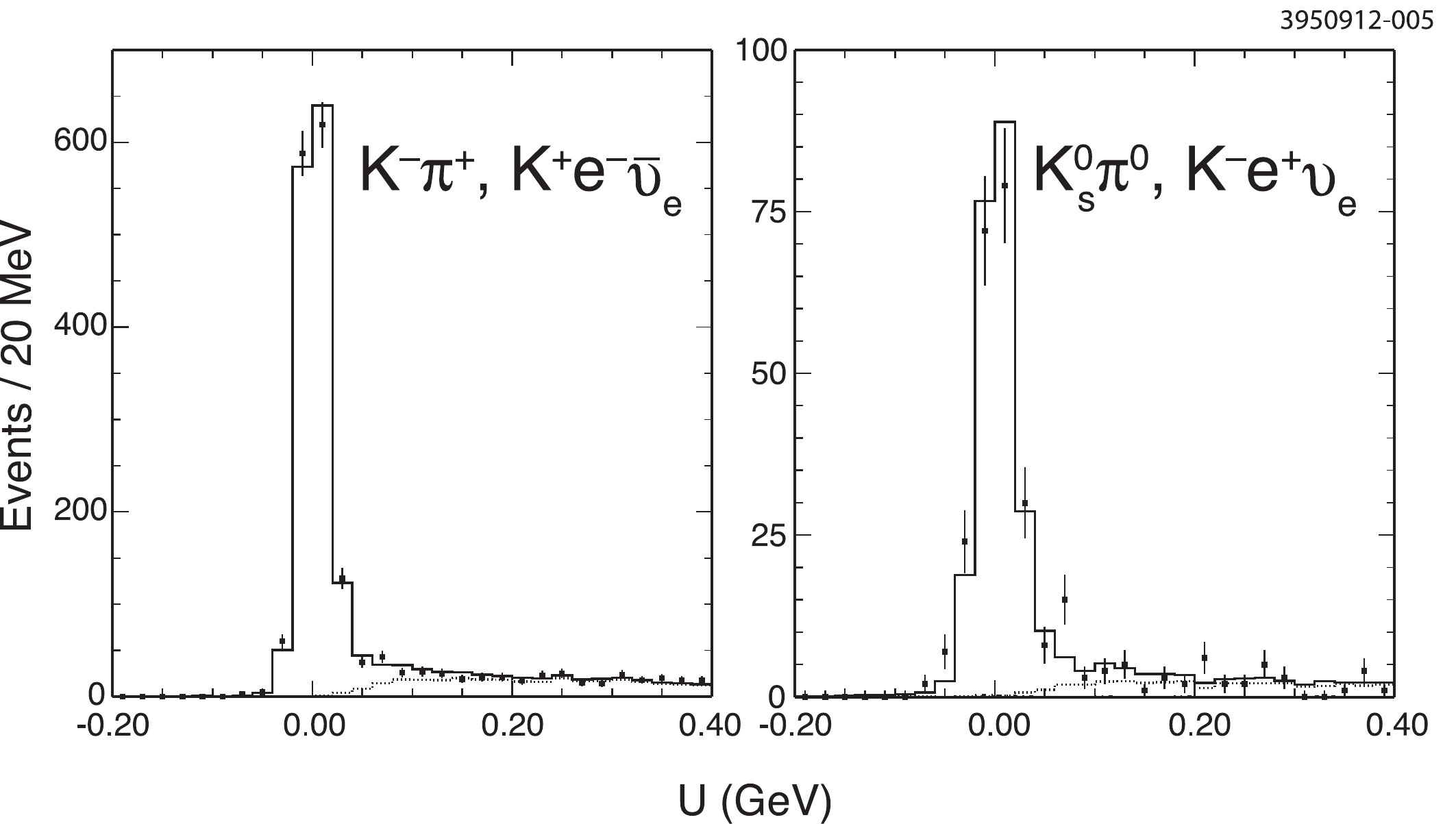}
\caption{Distributions of $U$ for $\{ K^-\pi^+, K^+e^-\bar\nu_e\}$ and
$\{K^0_S\pi^0, K^-e^+\nu_e\}$.  Data are shown as points with error bars.
The solid lines show
the total fits, and the dashed lines show the background components.}
\label{fig:kenu}
\end{figure}

\begin{figure}[htb]
\includegraphics*[width=\linewidth]{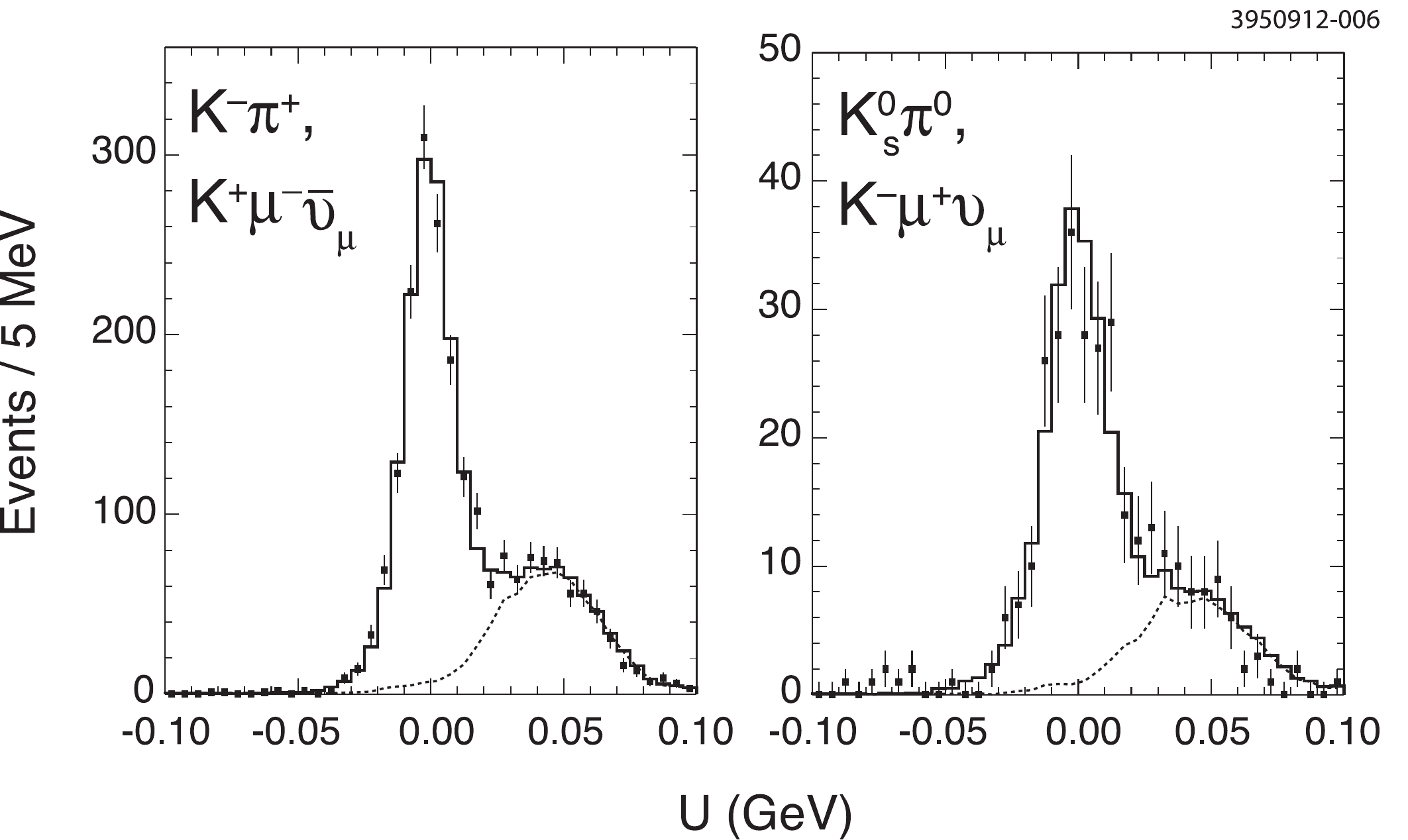}
\caption{Distributions of $U$ for $\{ K^-\pi^+, K^+\mu^-\bar\nu_\mu\}$ and
$\{K^0_S\pi^0, K^-\mu^+\nu_\mu\}$.  Data are shown as points with error bars.
The solid lines show
the total fits, and the dashed lines show the background components.}
\label{fig:kmunu}
\end{figure}

In our previous analysis of Ref.~\cite{tqca1}, we reconstructed
semielectronic final states inclusively, by identifying only the
electron and not the accompanying neutrino or hadronic system.
Also, we did not reconstruct semimuonic $D^0$ decays in Ref.~\cite{tqca1}.
In the current analysis, we replace inclusive reconstruction by
exclusive reconstruction of $K^- e^+\nu_e$ and $K^+ e^-\bar\nu_e$,
which allows us, in general,
to reduce systematic uncertainties because of lower background while
keeping roughly the same statistical power.
We also reconstruct $K^-\mu^+\nu_\mu$ and $K^+\mu^-\bar\nu_\mu$
without using the CLEO muon chambers because they are
insensitive to muons in the momentum range of interest
($p_\mu < 1$ GeV/$c$).
For both $Ke\nu_e$ and $K\mu\nu_\mu$, we begin with a fully reconstructed
hadronic ST candidate, selected as described in Section~\ref{sec:st},
with additional requirements on $M$ given in Secs.~\ref{sec:kenu}
and~\ref{sec:kmunu} below.
Each ST candidate is then combined with a kaon and lepton candidate
with opposite charges.
To extract the signal yields, we calculate the quantity
$U\equiv E_{\rm miss} - c p_{\rm miss}$, where
$E_{\rm miss}$ and $p_{\rm miss}$ are the missing energy
and the magnitude of the missing momentum, respectively, recoiling
against the observed
signal candidate particles in each event. Signal events
peak at $U=0$ GeV because of the undetected neutrino.
Representative $U$ distributions for $Ke\nu_e$ and $K\mu\nu_\mu$
are shown in Figs.~\ref{fig:kenu} and~\ref{fig:kmunu}.

\begin{figure}[htb]
\includegraphics*[width=\linewidth]{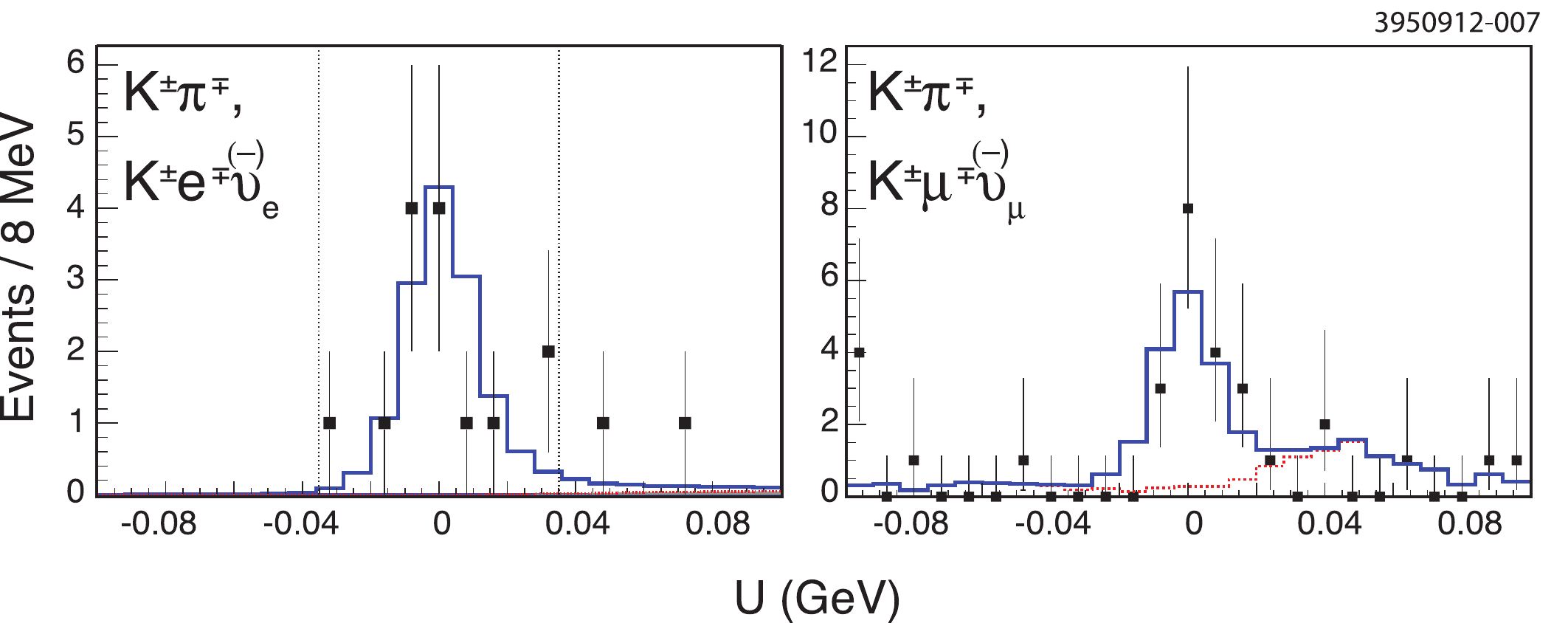}
\caption{Summed $U$ distributions for
$\{ K^-\pi^+, K^-e^+\nu_e\}$ and $\{ K^+\pi^-, K^+e^-\bar\nu_e\}$, as well
as for $\{ K^-\pi^+, K^-\mu^+\nu_\mu\}$ and $\{ K^+\pi^-, K^+\mu^-\bar\nu_\mu\}$.
Data are shown as points with error bars. For $Ke\nu_e$, the vertical lines mark
the signal region, the solid histogram shows the simulated signal distribution
normalized within the signal region to the number of observed events, and
the small dashed histogram at large $U$ values is the predicted background
from simulation.
For $K\mu\nu_\mu$, the solid histogram shows
the total fit, and the dashed histogram shows the background component.}
\label{fig:ws}
\end{figure}

When the ST mode is $K^\mp\pi^\pm$, we form both the Cabibbo-favored (CF)
combinations $\{K^-\pi^+, K^+\ell^-\bar\nu_\ell\}$ and
$\{K^+\pi^-, K^-\ell^+\nu_\ell\}$ as well as the doubly
Cabibbo-suppressed (DCS) combinations $\{K^-\pi^+, K^-\ell^+\nu_\ell\}$
and $\{K^+\pi^-, K^+\ell^-\bar\nu_\ell\}$. As shown in
Section~\ref{sec:formalism}, the ratio of CF and DCS yields
gives a direct measurement of the squared amplitude ratio $r^2$.
For the DCS combinations, we place an additional requirement
on the polar angle of the kaon in the $K^\mp\pi^\pm$ candidate of
$\left|\cos\theta\right| < 0.8$. This requirement selects only kaons in the
acceptance of the Ring Imaging \v{C}erenkov counter (RICH), and it
reduces the otherwise dominant
background from misidentified Cabibbo-favored combinations
to a negligible level.
Figure~\ref{fig:ws} shows the summed $U$ distributions for the
two DCS $Ke\nu_e$ modes and the two DCS $K\mu\nu_\mu$ modes.

For $K\ell\nu_\ell$ tagged with $K^0_S\pi^+\pi^-$, we consider sums of CF
combinations
$\{\bar Y_i, K^-\ell^+\nu_\ell\}$ and $\{Y_i, K^+\ell^-\bar\nu_\ell\}$, as well
as sums of Cabibbo-suppressed (CS) combinations
$\{Y_i, K^-\ell^+\nu_\ell\}$ and $\{\bar Y_i, K^+\ell^-\bar\nu_\ell\}$.
We include the CF and CS $\{Y_i, Ke\nu_e\}$ yields measured
in Ref.~\cite{kspipi}, and we perform a similar determination of
CF and CS $\{Y_i, K\mu\nu_\mu\}$ yields where we
select $K^0_S\pi^+\pi^-$ tags using the same criteria as in
Ref.~\cite{kspipi}.
In analogy with $K^\mp\pi^\pm$ tags, the ratios of CF and CS yields probes the
amplitude ratios $\rho_i^2$, integrated over each phase space bin.

\subsubsection{$K e \nu_e$}\label{sec:kenu}

For $Ke\nu_e$ modes, we require the fully reconstructed ST to have
$1.8530 \ {\rm GeV}/c^2 < M < 1.8780 \ {\rm GeV}/c^2$ for $K^0_S\pi^0$
and $K^0_S\pi^0\pi^0$,
$1.86 \ {\rm GeV}/c^2 < M < 1.87 \ {\rm GeV}/c^2$ for the DCS modes,
and $1.8585 \ {\rm GeV}/c^2 < M < 1.8775 \ {\rm GeV}/c^2$ for all other modes.
We identify electron candidates with the
same criteria as in Ref.~\cite{tqca1}, via a multivariate
discriminant that combines information from the ratio of the
energy deposited in the calorimeter to the measured
track momentum ($E/p$), ionization energy loss in the tracking chambers
($dE/dx$), and the RICH.
We fit the $U$ distributions to signal and background probability
distribution functions (PDFs) derived from simulated events and from
sidebands in $M$ and $\Delta E$ for the hadronic tag.
For DCS modes, because of the low background,
we simply count the number of events in the
signal region $|U| < 0.352$ GeV, and we estimate a relative
background contribution of ${\cal O}(10^{-3})$ from simulated events.

Table~\ref{tab:KenuYieldEffs} gives the semileptonic DT yields and efficiencies
for modes without $K^0_S\pi^+\pi^-$.
The uncorrelated systematic uncertainties are determined
from yield excursions under variation of the signal and background shapes,
histogram binning, and fit ranges. For DCS yields, we vary the size of the
signal region by one-third, and we assign additional systematic uncertainties
to the kaon polar-angle requirement and electron identification.
Yields for $Ke\nu_e$ tagged with $K^0_S\pi^+\pi^-$ (in 8 phase bins)
are taken from Ref.~\cite{kspipi}.

\begin{table}[htbp]
\begin{center}
\caption{DT yields and efficiencies including constituent branching fractions,
for $Ke\nu_e$ modes without $K^0_S\pi^+\pi^-$.
Yield uncertainties are statistical and uncorrelated systematic, respectively,
and efficiency uncertainties are statistical only.}
\label{tab:KenuYieldEffs}
\begin{tabular}{lcc}
\hline\hline
Mode & Yield & Efficiency (\%) \\
\hline
$K^-\pi^+, K^+e^-\bar\nu_e$ &
   $1523\pm 40\pm 16$ &
   $37.9\pm 0.2$ \\
$K^+\pi^-, K^+e^-\bar\nu_e$ &
   $5.0\pm 2.2\pm 0.9$ &
   $30.6\pm 0.2$ \\
$K^-K^+, K^+e^-\bar\nu_e$ &
   $156\pm 13\pm 4$ &
   $33.0\pm 0.5$ \\
$\pi^-\pi^+, K^+e^-\bar\nu_e$ &
   $70\pm 9\pm 5$ &
   $42.3\pm 0.9$ \\
$K^0_S\pi^0\pi^0, K^+e^-\bar\nu_e$ &
   $97\pm 11\pm 6$ &
   $10.6\pm 0.2$ \\
$K^0_S\pi^0, K^+e^-\bar\nu_e$ &
   $245\pm 16\pm 15$ &
   $20.1\pm 0.3$ \\
$K^0_S\eta, K^+e^-\bar\nu_e$ &
   $60\pm 8\pm 8$ &
   $6.7\pm 0.3$ \\
$K^0_S\omega, K^+e^-\bar\nu_e$ &
   $76\pm 11\pm 9$ &
   $7.9\pm 0.2$ \\
$K^-\pi^+, K^-e^+\nu_e$ &
   $9.0\pm 3.0\pm 1.6$ &
   $29.8\pm 0.2$ \\
$K^+\pi^-, K^-e^+\nu_e$ &
   $1603\pm 42\pm 23$ &
   $38.0\pm 0.2$ \\
$K^-K^+, K^-e^+\nu_e$ &
   $175\pm 14\pm 8$ &
   $33.6\pm 0.5$ \\
$\pi^-\pi^+, K^-e^+\nu_e$ &
   $64\pm 8\pm 1$ &
   $42.2\pm 0.9$ \\
$K^0_S\pi^0\pi^0, K^-e^+\nu_e$ &
   $108\pm 12\pm 6$ &
   $9.8\pm 0.2$ \\
$K^0_S\pi^0, K^-e^+\nu_e$ &
   $244\pm 16\pm 5$ &
   $20.2\pm 0.3$ \\
$K^0_S\eta, K^-e^+\nu_e$ &
   $35\pm 6\pm 2$ &
   $6.9\pm 0.3$ \\
$K^0_S\omega, K^-e^+\nu_e$ &
   $73\pm 10\pm 8$ &
   $7.5\pm 0.2$ \\
\hline\hline
\end{tabular}
\end{center}
\end{table}

\subsubsection{$K \mu \nu_\mu$}\label{sec:kmunu}

For all $K\mu\nu_\mu$ modes, we require
$1.86 \ {\rm GeV}/c^2 < M < 1.87 \ {\rm GeV}/c^2$
for the fully reconstructed ST.
We select muon candidates using the same criteria as for charged pion
candidates, except the particle identification requirements (on $dE/dx$ and
RICH) are applied to the muon mass hypothesis instead of the pion mass
hypothesis.
In reconstructing $K\mu\nu_\mu$, we reduce contamination from $Ke\nu_e$
by requiring the muon momentum to be
greater than 220 MeV/$c$, and we veto muon and kaon candiates that
also satisfy the multivariate electron discriminant described in
Section~\ref{sec:kenu}.
Requiring $p_{\rm miss} > 100$ MeV/$c$ suppresses
the $D\to K\pi$ background, for which $p_{\rm miss}$ peaks near zero.
Finally, we reject events that contain an additional shower with energy
greater than 100 MeV, in order to reduce the $K^-\pi^+\pi^0$ background
contribution. After this requirement, $K^-\pi^+\pi^0$ remains the dominant
background, but it is kinematically separated in $U$ from
signal $K^-\mu^+\nu_\mu$ because of both the
$\nu_\mu$-$\pi^0$ mass difference and the
$\mu^+$-$\pi^+$ mass difference.

For all $K\mu\nu_\mu$ modes, including the DCS modes, we determine
signal yields by fitting the $U$ distributions
to signal and background PDFs derived from simulated events.
For the non-DCS modes, the background PDFs are smoothed to reduce
the effect of statistical fluctuations in the simulated histograms.
Tables~\ref{tab:KmunuYieldEffs} and~\ref{tab:KmunuYieldEffsKspipi} give
the semileptonic DT yields and efficiencies for modes
with and without $K^0_S\pi^+\pi^-$, respectively.
The uncorrelated systematic uncertainties are determined
from yield excursions under variation of fit variations.
For DCS yields, we also include systematic uncertainties for the
kaon polar-angle requirement and the electron veto.

\begin{table}[htbp]
\begin{center}
\caption{DT yields and efficiencies including constituent branching fractions,
for $K\mu\nu_\mu$ modes without $K^0_S\pi^+\pi^-$.
Yield uncertainties are statistical and uncorrelated systematic, respectively,
and efficiency uncertainties are statistical only.}
\label{tab:KmunuYieldEffs}
\begin{tabular}{lcc}
\hline\hline
Mode & Yield & Efficiency (\%) \\
\hline
$K^-\pi^+, K^+\mu^-\bar\nu_\mu$ &
   $1442\pm 40\pm 13$ &
   $37.3\pm 0.2$ \\
$K^+\pi^-, K^+\mu^-\bar\nu_\mu$ &
   $7.0\pm 2.7\pm 1.1$ &
   $34.8\pm 0.2$ \\
$K^-K^+, K^+\mu^-\bar\nu_\mu$ &
   $121\pm 12\pm 0$ &
   $32.9\pm 0.2$ \\
$\pi^-\pi^+, K^+\mu^-\bar\nu_\mu$ &
   $63.3\pm 8.5\pm 1.0$ &
   $42.7\pm 0.2$ \\
$K^0_S\pi^0\pi^0, K^+\mu^-\bar\nu_\mu$ &
   $85.2\pm 10.6\pm 4.7$ &
   $8.6\pm 0.1$ \\
$K^0_S\pi^0, K^+\mu^-\bar\nu_\mu$ &
   $216\pm 16\pm 6$ &
   $18.3\pm 0.1$ \\
$K^0_S\eta, K^+\mu^-\bar\nu_\mu$ &
   $37.7\pm 6.4\pm 0.2$ &
   $6.5\pm 0.1$ \\
$K^0_S\omega, K^+\mu^-\bar\nu_\mu$ &
   $91.9\pm 10.5\pm 1.6$ &
   $7.1\pm 0.1$ \\
$K^-\pi^+, K^-\mu^+\nu_\mu$ &
   $9.8\pm 3.5\pm 1.6$ &
   $33.8\pm 0.2$ \\
$K^+\pi^-, K^-\mu^+\nu_\mu$ &
   $1446\pm 41\pm 13$ &
   $38.0\pm 0.2$ \\
$K^-K^+, K^-\mu^+\nu_\mu$ &
   $175\pm 14\pm 0$ &
   $32.5\pm 0.2$ \\
$\pi^-\pi^+, K^-\mu^+\nu_\mu$ &
   $74.5\pm 9.0\pm 1.2$ &
   $41.7\pm 0.2$ \\
$K^0_S\pi^0\pi^0, K^-\mu^+\nu_\mu$ &
   $88.0\pm 10.5\pm 4.8$ &
   $8.5\pm 0.1$ \\
$K^0_S\pi^0, K^-\mu^+\nu_\mu$ &
   $223\pm 16\pm 6$ &
   $18.1\pm 0.1$ \\
$K^0_S\eta, K^-\mu^+\nu_\mu$ &
   $33.0\pm 6.2\pm 0.2$ &
   $6.5\pm 0.1$ \\
$K^0_S\omega, K^-\mu^+\nu_\mu$ &
   $79.8\pm 10.3\pm 1.4$ &
   $7.2\pm 0.1$ \\
\hline\hline
\end{tabular}
\end{center}
\end{table}

\begin{table}[htbp]
\begin{center}
\caption{DT yields and efficiencies including constituent branching fractions,
for $K\mu\nu_\mu$ modes
with $K^0_S\pi^+\pi^-$. CF refers to the sum of Cabibbo-favored combinations
$\{\bar Y_i, K^-\mu^+\nu_\mu\}$ and $\{Y_i, K^+\mu^-\bar\nu_\mu\}$. Similarly,
CS refers to the sum of Cabibbo-suppressed combinations
$\{Y_i, K^-\mu^+\nu_\mu\}$ and $\{\bar Y_i, K^+\mu^-\bar\nu_\mu\}$.
Yield uncertainties are statistical and uncorrelated systematic, respectively,
and efficiency uncertainties are statistical only.}
\label{tab:KmunuYieldEffsKspipi}
\begin{tabular}{lcc}
\hline\hline
Mode & Yield & Efficiency (\%) \\
\hline
$Y_0, K\mu\nu_\mu$ (CF) &
   $162\pm 14\pm 2$ &
   $18.2\pm 0.2$ \\
$Y_1, K\mu\nu_\mu$ (CF) &
   $75.7\pm 9.3\pm 0.8$ &
   $18.4\pm 0.4$ \\
$Y_2, K\mu\nu_\mu$ (CF) &
   $132\pm 13\pm 1$ &
   $18.6\pm 0.3$ \\
$Y_3, K\mu\nu_\mu$ (CF) &
   $36.3\pm 6.4\pm 0.4$ &
   $18.9\pm 0.5$ \\
$Y_4, K\mu\nu_\mu$ (CF) &
   $67.7\pm 8.8\pm 0.7$ &
   $18.9\pm 0.3$ \\
$Y_5, K\mu\nu_\mu$ (CF) &
   $92.3\pm 10.4\pm 0.9$ &
   $17.8\pm 0.3$ \\
$Y_6, K\mu\nu_\mu$ (CF) &
   $120\pm 12\pm 1$ &
   $17.5\pm 0.3$ \\
$Y_7, K\mu\nu_\mu$ (CF) &
   $144\pm 13\pm 1$ &
   $17.8\pm 0.3$ \\
$Y_0, K\mu\nu_\mu$ (CS) &
   $66.0\pm 8.5\pm 0.7$ &
   $17.8\pm 0.4$ \\
$Y_1, K\mu\nu_\mu$ (CS) &
   $13.2\pm 4.1\pm 0.1$ &
   $17.9\pm 0.6$ \\
$Y_2, K\mu\nu_\mu$ (CS) &
   $33.0\pm 6.2\pm 0.3$ &
   $20.9\pm 0.8$ \\
$Y_3, K\mu\nu_\mu$ (CS) &
   $22.7\pm 4.9\pm 0.2$ &
   $20.0\pm 1.1$ \\
$Y_4, K\mu\nu_\mu$ (CS) &
   $46.8\pm 7.3\pm 0.5$ &
   $18.6\pm 0.6$ \\
$Y_5, K\mu\nu_\mu$ (CS) &
   $26.1\pm 5.7\pm 0.3$ &
   $18.8\pm 0.7$ \\
$Y_6, K\mu\nu_\mu$ (CS) &
   $21.1\pm 5.0\pm 0.2$ &
   $17.0\pm 0.7$ \\
$Y_7, K\mu\nu_\mu$ (CS) &
   $58.1\pm 8.2\pm 0.6$ &
   $17.7\pm 0.4$ \\
\hline\hline
\end{tabular}
\end{center}
\end{table}

\subsection{\boldmath Events with $Ke\nu_e$ and $K^0_L\pi^0$}\label{sec:twomissingparticles}

To reconstruct $\{Ke\nu_e, K^0_L\pi^0\}$, where both the $\nu_e$ and
$K^0_L$ are undetected, we adopt the technique described in
Refs.~\cite{Brower:1997be} and~\cite{Hokuue:2006nr} for identifying
events with two missing particles.
Knowing the energy and momentum magnitude of the two $D$ mesons in
the inital state, and having reconstructed the $K^\pm$, $e^\mp$, and
$\pi^0$ in the signal process, the direction of each $D$ meson
in the $e^+e^-$ center of mass frame
can be constrained to a cone around the flight direction of
the $K^\pm e^\mp$ system or the $\pi^0$. In signal events,
the $D^0$ and $\bar D^0$ are collinear (within detector resolution),
so when one of the cones is reflected through the origin, the two
cones intersect. Background events typically have non-intersecting
cones.

Following Ref.~\cite{Hokuue:2006nr}, we calculate the quantity
\begin{equation}
x_D^2 \equiv 1 - \frac{1}{\sin^2\theta_{Ke,\pi^0}}
\left(
\cos^2\theta_{D,Ke} + \cos^2\theta_{D,\pi^0} +
2 \cos\theta_{Ke,\pi^0} \cos\theta_{D,Ke} \cos\theta_{D,\pi^0}
\right),
\end{equation}
where $\theta_{Ke,\pi^0}$ is the angle between the $K^\pm e^\mp$ system
and the $\pi^0$ candidate, and $\theta_{D,Ke}$ and $\theta_{D,\pi^0}$
are the opening half-angles of the $D$ cones around the $K^\pm e^\mp$
system and the $\pi^0$, respectively. By construction, $x_D^2$ is less
than or equal to 1,
and when the cones do not intersect, then $x_D^2 < 0$.
Signal events lie mostly in the range
$0 \leq x_D^2 \leq 1$, with a small tail extending to $x_D^2 < 0$
due to mismeasurement.

In addition to the previously described criteria for $K^\pm$,
$e^\mp$, and $\pi^0$ candidates, we recover electron bremsstrahlung
photons by augmenting the electron four-momentum by any showers
located within 100 mrad of the track direction and unassigned
to any other particle. Also, kaons that satisfy the electron
identification requirements are rejected.

We apply two additional requirements to suppress the background from
radiative
$e^+e^-$ Bhabha scattering events: the electron momentum is required to be
less than 1 GeV/$c$, and the $\pi^0$ candidate must not have daughter showers
that lie within 100 mrad of the electron candidate.
To reduce the background from $\{Ke\nu_e, K^0_S\pi^0\}$ events, we reject
events with extra tracks or $\pi^0/\eta$ candidates.

\begin{figure}[htb]
\includegraphics*[width=\linewidth]{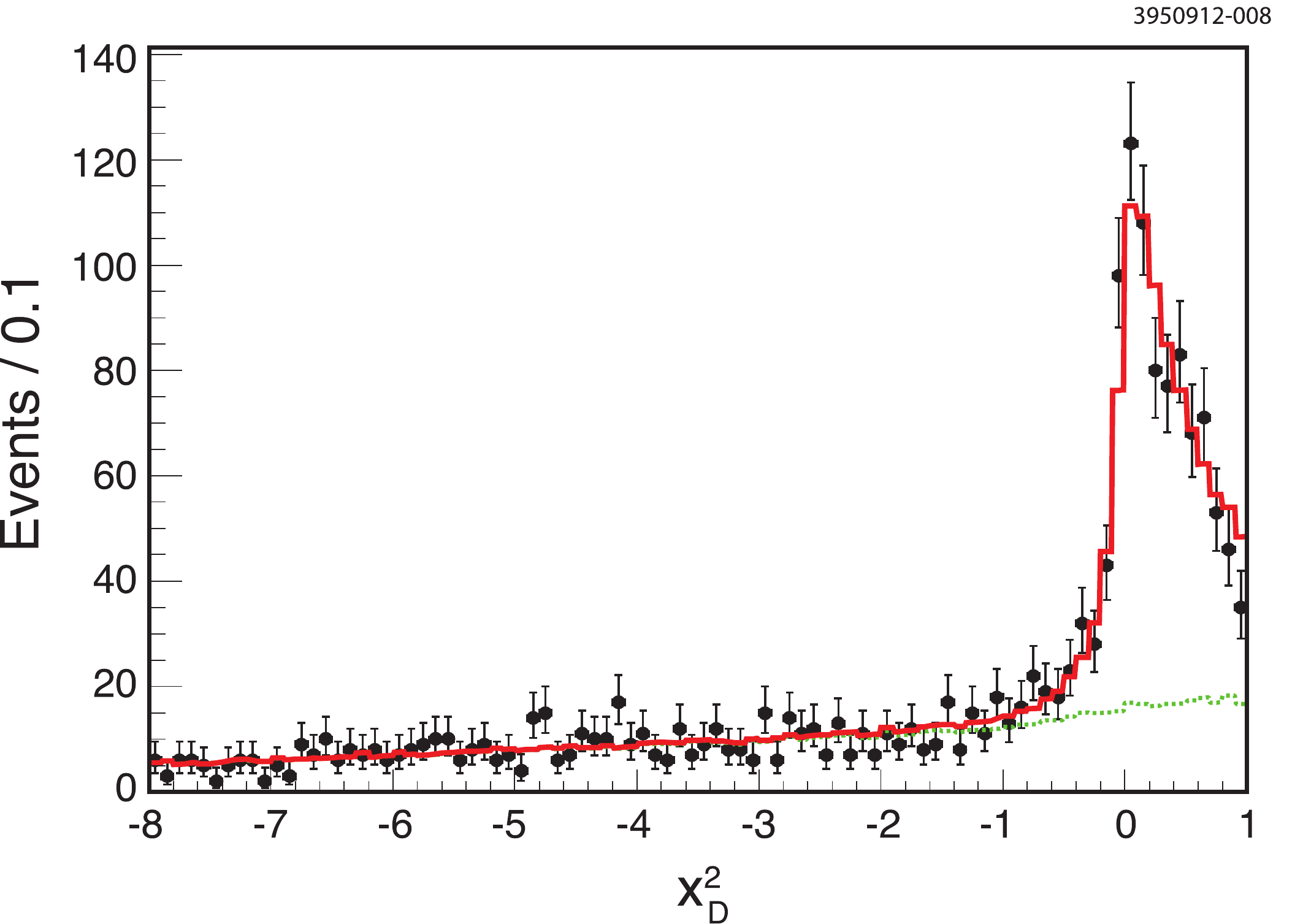}
\caption{Distribution of $x_D^2$ for $\{Ke\nu_e, K^0_L\pi^0\}$.
Data are shown as points with error bars.
The solid line shows
the total fit, and the dashed line shows the sum of background components.}
\label{fig:xD2}
\end{figure}

Figure~\ref{fig:xD2} shows the $x_D^2$ distribution, which is fitted
to signal and background shapes from simulated events. The fitted
yield is $764\pm 36\pm 23$, where the uncorrelated systematic uncertainty 
receives dominant contributions from the radiative Bhabha background
determination, variations in background and signal shapes, and uncertainties
in the modeling of extra tracks and
$\pi^0/\eta$ candidates. The signal efficiency determined from simulated
events is $(34.58\pm 0.04)\%$.

\subsection{Crossfeed and Peaking Backgrounds}~\label{sec:backgrounds}

As in Ref.~\cite{tqca1}, crossfeed among signal modes and
peaking background contributions are subtracted by the fitter, using
crossfeed probabilities and background efficiencies determined
from simulated events, as well as
branching fractions for peaking background processes~\cite{pdg12}.
These inputs are listed in Tables~\ref{tab:crossfeedSources}
and~\ref{tab:backgroundSources}, and
we account for correlated uncertainties among them.
Backgrounds in DT modes are assumed to contribute at the same rate as
for ST modes, with corrections for
quantum correlation effects. When a ST background process does not
exactly match the particle content of the corresponding signal process,
then that background does not contribute to DT modes because of the
additional constriaints imposed by DT reconstruction.
In general, peaking backgrounds contribute less than 1\% to the measured
yields, except in $K^0_L\pi^0$ modes (1--2\%), $K^0_L\eta$ modes (2--5\%),
$\{Ke\nu_e, K^0_L\pi^0\}$ (1\%),
ST $K^0_S\pi^0\pi^0$ (3\%),
Cabibbo-favored $\{K^0_S\pi^+\pi^-, K^0_S\pi^+\pi^-\}$ (1--19\%),
and Cabibbo-suppressed $\{K^0_S\pi^+\pi^-, K^0_S\pi^+\pi^-\}$ (2--56\%).
The large background fractions in the last two categories occur in modes
with small numbers of observed events.

\begin{table}[htbp]
\begin{center}
\caption{Crossfeed probabilities among signal modes. Ranges are given
for groups of modes.}
\label{tab:crossfeedSources}
\begin{tabular}{rclc}
\hline\hline
Crossfeed &$\to$& Signal & Efficiency (\%) \\
\hline
$K^+\pi^-$ &$\to$& $K^-\pi^+$ & $0.088\pm 0.002$ \\
$K^-\pi^+$ &$\to$& $K^+\pi^-$ & $0.089\pm 0.002$ \\
$K^0_S\omega$ &$\to$& $K^0_S\pi^0\pi^0$ & $0.080\pm 0.003$ \\
$K^-\pi^+$, $K^0_S\pi^0$ &$\to$& $K^-\pi^+$, $K^0_L\pi^0$ & $0.45\pm 0.02$ \\
$K^+\pi^-$, $K^0_S\pi^0$ &$\to$& $K^+\pi^-$, $K^0_L\pi^0$ & $0.43\pm 0.02$ \\
$K^-\pi^+$, $K^0_S\eta$ &$\to$& $K^-\pi^+$, $K^0_L\eta$ & $0.13\pm 0.01$ \\
$K^+\pi^-$, $K^0_S\eta$ &$\to$& $K^+\pi^-$, $K^0_L\eta$ & $0.12\pm 0.01$ \\
$K^-\pi^+$, $K^0_L\pi^0$ &$\to$& $K^-\pi^+$, $K^0_L\eta$ & $0.14\pm 0.01$ \\
$K^+\pi^-$, $K^0_L\pi^0$ &$\to$& $K^+\pi^-$, $K^0_L\eta$ & $0.13\pm 0.01$ \\
$K^0_S\pi^0$, $K^0_L\pi^0$ &$\to$& $K^0_S\pi^0$, $K^0_L\eta$ & $0.06\pm 0.01$ \\
$K^0_S\eta$, $K^0_L\pi^0$ &$\to$& $K^0_S\eta$, $K^0_L\eta$ & $0.04\pm 0.01$ \\
$K^0_S\omega$, $K^0_L\pi^0$ &$\to$& $K^0_S\omega$, $K^0_L\eta$ & $0.03\pm 0.01$ \\
$K^-\pi^+$, $K^0_S\omega$ &$\to$& $K^-\pi^+$, $K^0_L\omega$ & $0.13\pm 0.01$ \\
$K^+\pi^-$, $K^0_S\omega$ &$\to$& $K^+\pi^-$, $K^0_L\omega$ & $0.12\pm 0.01$ \\
$K^-\pi^+$, $K^0_S\pi^0\pi^0$ &$\to$& $K^-\pi^+$, $K^0_L\pi^0\pi^0$ & $0.47\pm 0.02$ \\
$K^+\pi^-$, $K^0_S\pi^0\pi^0$ &$\to$& $K^+\pi^-$, $K^0_L\pi^0\pi^0$ & $0.46\pm 0.02$ \\
$K^-\pi^+$, $K^0_S\pi^0$ &$\to$& $K^-\pi^+$, $K^0_L\pi^0\pi^0$ & $0.61\pm 0.03$ \\
$K^+\pi^-$, $K^0_S\pi^0$ &$\to$& $K^+\pi^-$, $K^0_L\pi^0\pi^0$ & $0.62\pm 0.03$ \\
$K^+K^-$, $K^0_S\pi^0$ &$\to$& $K^+K^-$, $K^0_L\pi^0\pi^0$ & $0.49\pm 0.02$ \\
$\pi^+\pi^-$, $K^0_S\pi^0$ &$\to$& $\pi^+\pi^-$, $K^0_L\pi^0\pi^0$ & $0.68\pm 0.03$ \\
$K^0_S\pi^0\pi^0$, $K^0_S\pi^0$ &$\to$& $K^0_S\pi^0\pi^0$, $K^0_L\pi^0\pi^0$ & $0.16\pm 0.01$ \\
$K^-\pi^+$, $K^0_L\pi^0$ &$\to$& $K^-\pi^+$, $K^0_L\pi^0\pi^0$ & $0.11\pm 0.01$ \\
$K^+\pi^-$, $K^0_L\pi^0$ &$\to$& $K^+\pi^-$, $K^0_L\pi^0\pi^0$ & $0.12\pm 0.01$ \\
$Ke\nu_e$, $K^0_S\pi^0$ &$\to$& $Ke\nu_e$, $K^0_L\pi^0$ & $2.31\pm 0.03$ \\
$K^-\pi^+$, $K^+e^-\bar\nu_e$ &$\to$& $K^-\pi^+$, $K^-e^+\nu_e$ & $0.0048\pm 0.0022$ \\
$K^+\pi^-$, $K^-e^+\nu_e$ &$\to$& $K^-\pi^+$, $K^-e^+\nu_e$ & $0.0057\pm 0.0025$ \\
$K^-\pi^+$, $K^+e^-\bar\nu_e$ &$\to$& $K^+\pi^-$, $K^+e^-\bar\nu_e$ & $0.0038\pm 0.0020$ \\
$K^+\pi^-$, $K^-e^+\nu_e$ &$\to$& $K^+\pi^-$, $K^+e^-\bar\nu_e$ & $0.0019\pm 0.0014$ \\
$K^-\pi^+$, $K^+\mu^-\bar\nu_\mu$ &$\to$& $K^-\pi^+$, $K^-\mu^+\nu_\mu$ & $0.0029\pm 0.0017$ \\
$K^+\pi^-$, $K^-\mu^+\nu_\mu$ &$\to$& $K^-\pi^+$, $K^-\mu^+\nu_\mu$ & $0.0067\pm 0.0026$ \\
$K^-\pi^+$, $K^+\mu^-\bar\nu_\mu$ &$\to$& $K^+\pi^-$, $K^+\mu^-\bar\nu_\mu$ & $0.0019\pm 0.0014$ \\
$K^+\pi^-$, $K^-\mu^+\nu_\mu$ &$\to$& $K^+\pi^-$, $K^+\mu^-\bar\nu_\mu$ & $0.0010\pm 0.0010$ \\
$K^0_S\eta$, $K\mu\nu_\mu$ &$\to$& $K^0_S\pi^+\pi^-$, $K\mu\nu_\mu$ & $(0$--$3)\times 10^{-3}$ \\
$K^0_S\omega$, $K\mu\nu_\mu$ &$\to$& $K^0_S\pi^+\pi^-$, $K\mu\nu_\mu$ & $(0$--$5)\times 10^{-2}$ \\
\hline\hline
\end{tabular}
\end{center}
\end{table}

\begin{table}[htbp]
\begin{center}
\caption{Peaking background branching fractions and efficiencies including
constituent branching fractions. Ranges are given for groups of modes.
Fully reconstructed hadronic tag modes are denoted by $X$.
Backgrounds marked by an asterisk (*) occur only in STs, not DTs.}
\label{tab:backgroundSources}
\begin{tabular}{rclcc}
\hline\hline
Background &$\to$& Signal & ${\cal B}_{\rm bkg}$ (\%)~\cite{pdg12} &
        Efficiency (\%) \\
\hline
$K^0_S\pi^+\pi^-$ &$\to$& $K^0_S\pi^0\pi^0$ &
$2.94\pm 0.16$ & $0.0076\pm 0.005$ \\
$K^-\pi^+\pi^0$ (*) &$\to$& $K^0_S\pi^0\pi^0$ &
$13.9\pm 0.5$ & $0.0027\pm 0.0001$ \\
$D^+\to K^0_S\pi^+\pi^0$ (*) &$\to$& $K^0_S\pi^0\pi^0$ &
$6.90\pm 0.32$ & $0.0594\pm 0.0010$ \\
$\rho^+\pi^-$ &$\to$& $K^0_S\pi^0$ &
$1.447\pm 0.046$ & $0.078\pm 0.004$ \\
$\rho^0\pi^0$ &$\to$& $K^0_S\pi^0$ &
$0.373\pm 0.022$ & $0.011\pm 0.004$ \\
Generic $D^0\bar D^0$ (*) &$\to$& $K^0_S\pi^0$ & 100 & $0.0006\pm 0.0001$ \\
Generic $D^+D^-$ (*) &$\to$& $K^0_S\pi^0$ & 100 & $0.0003\pm 0.0002$ \\
$\eta\pi^0$, $X$ &$\to$& $K^0_L\pi^0$, $X$ &
$0.064\pm 0.011$ & $0.3$--$1.4$ \\
$\pi^0\pi^0$, $X$ &$\to$& $K^0_L\pi^0$, $X$ &
$0.080\pm 0.008$ & $0.8$--$4.3$ \\
$\eta\pi^0$, $X$ &$\to$& $K^0_L\eta$, $X$ &
$0.064\pm 0.011$ & $0.1$--$0.3$ \\
$\eta\eta$, $X$ &$\to$& $K^0_L\eta$, $X$ &
$0.167\pm 0.019$ & $0.2$--$0.8$ \\
$Ke\nu_e$, $K^0_L\pi^0\pi^0$ &$\to$& $Ke\nu_e$, $K^0_L\pi^0$ &
From fitter & $0.45\pm 0.04$ \\
$K^0_L\pi^+\pi^-$, $K\mu\nu_\mu$ &$\to$& $K^0_S\pi^+\pi^-$, $K\mu\nu_\mu$ &
$2.94\pm 0.16$ & $(0$--$6)\times 10^{-3}$ \\
$\pi^+\pi^-\pi^+\pi^-$, $X$ &$\to$& $K^0_S\pi^+\pi^-$, $X$ &
$0.744\pm 0.021$ & $(1$--$25)\times 10^{-2}$ \\
\hline\hline
\end{tabular}
\end{center}
\end{table}


\section{External Measurements}\label{sec:externalMeas}

Unlike our previous analysis in Ref.~\cite{tqca1}, we do not include any
external branching fraction or $R_{\rm WS}$ measurements in the fit.
With our increased data sample,
branching fraction measurements from other experiments do not have a
significant impact on our precision, and our direct measurement of
$r^2$ (from DCS $\{K\pi, K\ell\nu_\ell\}$) obviates the need for
input on $r^2$ from $R_{\rm WS}$.
We do, however, include the external
measurements of mixing parameters shown in Table~\ref{tab:externalYYPrime},
where $x'^2 \equiv 2R_{\rm M} - y'^2$.
Averages from Ref.~\cite{hfag} in this Table do not include
previous CLEO-c results.
Correlation coefficients that are not shown in
Table~\ref{tab:externalYYPrime} are taken to be zero.
Also, since we neglect $CP$ violation, we assume $y_{CP} = y$.
In the fit, we use the signed
measurement of $x$ along with $y'$ to resolve the
sign ambiguity in $\sin\delta$ and the $s_i$,
even though we quote a fitted value only for $x^2$.

\begin{table}[htbp]
\begin{center}
\caption{External measurements of $y_{CP}$, $x$, $y$, $r^2$ $y'$ and $x'^2$.
Correlation coefficients $\rho$
are for the measurements in the same column.}
\label{tab:externalYYPrime}
\begin{tabular}{lccc}
\hline\hline
Parameter & \multicolumn{3}{c}{Value(s) (\%)} \\
\hline
$y_{CP}$ & \multicolumn{3}{c}{$1.064\pm 0.209$~\cite{hfag}} \\
$x$ & \multicolumn{3}{c}{$0.419\pm 0.211$~\cite{hfag}} \\
$y$ & \multicolumn{3}{c}{$0.456\pm 0.186$~\cite{hfag}} \\
$r^2$  & $0.364\pm 0.017$~\cite{kpiBelle}
  & $0.303\pm 0.016\pm 0.010$~\cite{kpiBABAR}
  & $0.304\pm 0.055$~\cite{kpiCDF} \\
$y'$   & $0.06^{+0.40}_{-0.39}$~\cite{kpiBelle}
   & $0.97\pm 0.44\pm 0.31$~\cite{kpiBABAR}
   & $0.85\pm 0.76$~\cite{kpiCDF} \\
$x'^2$ & $0.018^{+0.021}_{-0.023}$~\cite{kpiBelle}
 & $-0.022\pm 0.030\pm 0.021$~\cite{kpiBABAR}
 & $-0.012\pm 0.035$~\cite{kpiCDF} \\
$\rho(r^2, y')$ & $-83.4$~\cite{hfag2010}
                & $-87$~\cite{hfag2010}
                & $-97.1$~\cite{hfag2010} \\
$\rho(r^2, x'^2)$ & $+65.5$~\cite{hfag2010}
                  & $+77$~\cite{hfag2010}
                  & $+92.3$~\cite{hfag2010} \\
$\rho(y', x'^2)$ & $-90.9$~\cite{hfag2010}
                 & $-94$~\cite{hfag2010}
                 & $-98.4$~\cite{hfag2010} \\
\hline\hline
\end{tabular}
\end{center}
\end{table}


\section{Systematic Uncertainties}\label{sec:systematics}

We include systematic uncertainties directly in the fit. Uncorrelated
uncertainties for each yield are discussed above in Section~\ref{sec:selection}.
Correlated uncertainties are given in Tables~\ref{tab:particleSysts}
and~\ref{tab:modeSysts}. When the fit is performed without external
measurements, these correlated uncertainties cancel in all the fit
parameters except ${\cal N}$ and the branching fractions.
Uncertainties on the peaking background branching fractions in
Table~\ref{tab:backgroundSources} and the external measurements
in Table~\ref{tab:externalYYPrime} are also included as systematic
uncertainties.

In Table~\ref{tab:particleSysts}, we list the correlated systematic
uncertainties on reconstruction and particle identification efficiencies
for individual final state
particles. Those for $\pi^\pm$, $K^\pm$, $\pi^0$, $K^0_S$, and
$e^\pm$ reconstruction and particle identification (PID) are
determined using the techniques described in
Refs.~\cite{dhad281,dsl818}.
For $K^0_S$, we include additional contributions
for the flight significance and invariant mass requirements.
For $\eta$, we use the same uncertainties as in Ref.~\cite{tqca1}.
The $K^0_L$ uncertainties are determined by varying the selection
requirements.
The above studies also provide efficiency corrections, which are
included in the signal and background efficiencies listed in this paper.

\begin{table}[htbp]
\begin{center}
\caption{Correlated, fractional efficiency systematic uncertainties and
the schemes for applying them in the fit.}
\label{tab:particleSysts}
\begin{tabular}{lcc}
\hline\hline
Source & Uncertainty (\%) & Scheme \\ \hline
Track finding & 0.3 & per track \\
$K^\pm$ hadronic interactions & 0.5 & per $K^\pm$ \\
$K^0_S$ finding & 0.9 & per $K^0_S$ \\
$\pi^0$ finding & 2.0 & per $\pi^0$ \\
$\eta$ finding & 4.0 & per $\eta$ \\
$dE/dx$ and RICH & 0.1 & per $\pi^\pm$ PID cut \\
$dE/dx$ and RICH & 0.1 & per $K^\pm$ PID cut \\
Electron identification & 0.4 & per $e^\pm$ \\
$K^0_L$ shower veto & 0.4 & per $K^0_L$ \\
$K^0_L$ background subtraction & 0.7 & per $K^0_L$ \\
$K^0_L$ track veto & 0.3 & per $K^0_L$ \\
$K^0_L$ signal shape & 1.4 & per $K^0_L$ \\
\hline\hline
\end{tabular}
\end{center}
\end{table}

Table~\ref{tab:modeSysts} shows mode-dependent uncertainties, including
those for initial state radiation (ISR), final state radiation (FSR),
and selection requirements in two-track modes that veto radiative
Bhabhas and cosmic muons. Most of the uncertainties in Table~\ref{tab:modeSysts}
are determined with the same techniques as in Ref.~\cite{tqca1},
updated to the current data and simulated samples.
We assess uncertainties for new modes by 
relaxing selection requirements and noting the
resultant change in efficiency-corrected yield.

\begin{table}[htbp]
\begin{center}
\caption{Correlated, mode-dependent fractional systematic uncertainties
in percent for STs.  An asterisk (*) marks those uncertainties that
are correlated among modes.  The schemes by which these uncertainties are
applied to DTs are also given, along with the fractional DT
uncertainty, $\lambda_{\rm DT}$, on mode $\{A, B\}$ for ST uncertainties of
$\alpha$ on mode $A$ and $\beta$ on mode $B$.}
\label{tab:modeSysts}
\begin{tabular}{lcccccl}
\hline\hline
   & $\Delta E$ & ISR* & FSR* & Lepton Veto* & \multicolumn{2}{l}{Other} \\
\hline
$K^\mp\pi^\pm$ & 0.5 & 0.5 & 0.9 & 0.5 \\
$K^+K^-$ & 0.9 & 0.5 & 0.5 & 0.4 & 0.5 & $K^\pm$ $\cos\theta$ cut \\
$\pi^+\pi^-$ & 1.9 & 0.5 & 1.4 & 3.2 \\
$K^0_S\pi^0\pi^0$ & 2.6 & 0.5 & & & 1.5 & $K^0_S$ daughter PID \\
        & & & & & 0.7 & resonant substructure \\
$K^0_S\pi^0$ & 0.9 & 0.5 \\
$K^0_S\eta$ & 5.5 & 0.5 & & & 0.3 & $\eta$ mass cut \\
        & & & & & 0.7 & ${\cal B}(\eta\to\gamma\gamma)$~\cite{pdg12} \\
$K^0_S\omega$ & 1.2 & 0.5 & 0.6 & & 0.1 & $\omega$ mass cut/SB subt. \\
        & & & & & 0.8 & ${\cal B}(\omega\to\pi^+\pi^-\pi^0)$~\cite{pdg12} \\
$K^0_L\pi^0(\pi^0)$ & & 0.5 & & & & \\
$K^0_L\eta$ & & 0.5 & & & 1.6 & extra $\pi^0$ veto \\
$K^0_L\omega$ & & 0.5 & 0.6 & & 0.1 & $\omega$ mass cut/SB subt. \\
        & & & & & 0.8 & ${\cal B}(\omega\to\pi^+\pi^-\pi^0)$~\cite{pdg12} \\
$K^0_S\pi^+\pi^-$ & 0.9 & 0.5 & 1.4 \\
$Ke\nu_e$ & & 0.5 & 0.3 & & 2.0 & spectrum extrapolation \\
$K\mu\nu_\mu$ & & 0.5 & 0.3 & & 2.0 & spectrum extrapolation \\
        & & & & & 0.4 & extra shower veto \\
\hline
Scheme & per $D$ & per yield & per $D$ & per ST & \multicolumn{2}{l}{per $D$} \\
$\lambda_{\rm DT}$ & $\sqrt{\alpha^2+\beta^2}$ & $(\alpha+\beta)/2$
        & $\alpha+\beta$ & 0
        & \multicolumn{2}{l}{$\sqrt{\alpha^2+\beta^2}$} \\
\hline\hline
\end{tabular}
\end{center}
\end{table}


\section{Fit Results}\label{sec:results}

We combine the 261 yield measurements discussed above, along with estimates
of efficiencies and background contributions, in a $\chi^2$ fit that
determines 51 free parameters: ${\cal N}$; $y$; $r^2$; $\cos\delta$;
$\sin\delta$; $x^2$; $\rho_i^2$, $c_i$, and $s_i$ for each of the 8
phase bins in $K^0_S\pi^+\pi^-$; and 21 branching fractions.
We tested the analysis technique using a simulated sample of
quantum-correlated $D^0\bar D^0$ decays with an effective integrated
luminosity of 10 times our data sample.
In Tables~\ref{tab:results1} and~\ref{tab:results2}, we show the results
of two fits: one with no external inputs (Standard Fit) and one including the
measurements in Table~\ref{tab:externalYYPrime} (Extended Fit).
The fits are performed using both statistical and systematic uncertainties
on the input measurements. We evaluate statistical uncertainties on the fit
parameters by performing a second set of fits using only statistical
uncertainties on the inputs. Then, we compute the systematic uncertainty on
a given fit parameter by taking the quadrature difference between its total
uncertainty and its statistical uncertainty.
When a fit parameter is directly constrained by an external
measurement, we quote only one uncertainty.
Asymmetric uncertainties are determined from the likelihood scans discussed
below.
Table~\ref{tab:correlations} gives the correlation coefficients for
$y$, $r^2$, $\cos\delta$, $\sin\delta$, and $x^2$.
The full correlation matrices can be found on EPAPS~\cite{epaps}.

\begin{table}[htb]
\caption{Results from the Standard Fit and the Extended Fit for
all parameters except branching fractions.
Uncertainties are statistical and systematic, respectively.
In the Extended Fit, we quote only one uncertainty for $y$, $r^2$, and $x^2$,
which are directly constrained by an external measurement.}
\label{tab:results1}
\begin{tabular}{lcc}
\hline\hline
Parameter & ~~~~~~~Standard Fit~~~~~~~ & ~~~~~~~Extended Fit~~~~~~~ \\
\hline
${\cal N}$ $(10^6)$
  & $3.092\pm 0.050\pm 0.040$ & $3.114\pm 0.050\pm 0.040$ \\
$y$ (\%)
  & $4.2\pm 2.0\pm 1.0$ & $0.636\pm 0.114$ \\
$r^2$ $(\%)$
  & $0.533\pm 0.107\pm 0.045$ & $0.333\pm 0.008$ \\
$\cos\delta$
  & $0.81^{+0.22+0.07}_{-0.18-0.05}$ & $1.15^{+0.19+0.00}_{-0.17-0.08}$ \\
$\sin\delta$
  & $-0.01\pm 0.41 \pm 0.04$ & $0.56^{+0.32+0.21}_{-0.31-0.20}$ \\
$x^2$ $(\%)$
  & $0.06\pm 0.23\pm 0.11$ & $0.0022\pm 0.0023$ \\
$\rho_0^2$
  & $0.337\pm 0.030\pm 0.006$ & $0.352\pm 0.032\pm 0.005$ \\
$\rho_1^2$
  & $0.270\pm 0.044\pm 0.005$ & $0.280\pm 0.047\pm 0.000$ \\
$\rho_2^2$
  & $0.235\pm 0.028\pm 0.003$ & $0.252\pm 0.028\pm 0.004$ \\
$\rho_3^2$
  & $0.399\pm 0.066\pm 0.005$ & $0.416\pm 0.069\pm 0.000$ \\
$\rho_4^2$
  & $0.592\pm 0.067\pm 0.010$ & $0.623\pm 0.071\pm 0.000$ \\
$\rho_5^2$
  & $0.343\pm 0.044\pm 0.000$ & $0.329\pm 0.040\pm 0.008$ \\
$\rho_6^2$
  & $0.146\pm 0.023\pm 0.000$ & $0.145\pm 0.023\pm 0.000$ \\
$\rho_7^2$
  & $0.445\pm 0.039\pm 0.002$ & $0.439\pm 0.039\pm 0.003$ \\
$c_0$
  & $-0.76\pm 0.06\pm 0.01$ & $-0.73\pm 0.06\pm 0.01$ \\
$c_1$
  & $-0.75\pm 0.11\pm 0.00$ & $-0.72\pm 0.11\pm 0.02$ \\
$c_2$
  & $0.00\pm 0.10\pm 0.01$ & $0.03\pm 0.10\pm 0.02$ \\
$c_3$
  & $0.45\pm 0.15\pm 0.01$ & $0.47\pm 0.14\pm 0.01$ \\
$c_4$
  & $0.95\pm 0.07\pm 0.01$ & $0.95\pm 0.07\pm 0.00$ \\
$c_5$
  & $0.79\pm 0.09\pm 0.01$ & $0.81\pm 0.09\pm 0.00$ \\
$c_6$
  & $-0.20\pm 0.13\pm 0.02$ & $-0.16\pm 0.13\pm 0.01$ \\
$c_7$
  & $-0.41\pm 0.07\pm 0.01$ & $-0.39\pm 0.07\pm 0.01$ \\
$s_0$
  & $0.55\pm 0.16\pm 0.00$ & $0.61\pm 0.15\pm 0.02$ \\
$s_1$
  & $0.53\pm 0.28\pm 0.00$ & $0.56\pm 0.27\pm 0.03$ \\
$s_2$
  & $0.93\pm 0.15\pm 0.00$ & $0.91\pm 0.15\pm 0.02$ \\
$s_3$
  & $0.47\pm 0.30\pm 0.00$ & $0.52\pm 0.29\pm 0.01$ \\
$s_4$
  & $0.55\pm 0.24\pm 0.00$ & $0.60\pm 0.23\pm 0.02$ \\
$s_5$
  & $-0.71\pm 0.24\pm 0.00$ & $-0.69\pm 0.24\pm 0.00$ \\
$s_6$
  & $-0.42\pm 0.27\pm 0.06$ & $-0.17\pm 0.29\pm 0.03$ \\
$s_7$
  & $-0.30\pm 0.18\pm 0.04$ & $-0.21\pm 0.19\pm 0.03$ \\
\hline
$\chi^2_{\rm fit}$/ndof & 193.2/210 & 214.7/222\\
\hline\hline
\end{tabular}
\end{table}

\begin{table}[htb]
\caption{Branching fraction results from the Standard Fit and the Extended Fit.
Uncertainties are statistical and systematic, respectively.}
\label{tab:results2}
\begin{tabular}{lcc}
\hline\hline
Parameter & ~~~~~Standard Fit~~~~~ & ~~~~~Extended Fit~~~~~ \\
\hline
${\cal B}(K^-\pi^+)$ (\%) &
$3.77\pm 0.06\pm 0.05$ & $3.76\pm 0.06\pm 0.05$ \\
${\cal B}(K^-K^+)$ $(10^{-3})$ &
$3.99\pm 0.07\pm 0.08$ & $3.98\pm 0.07\pm 0.08$ \\
${\cal B}(\pi^-\pi^+)$ $(10^{-3})$ &
$1.36\pm 0.03\pm 0.04$ & $1.37\pm 0.03\pm 0.04$ \\
${\cal B}(K^0_S\pi^0\pi^0)$ (\%) &
$0.99\pm 0.02\pm 0.06$ & $0.99\pm 0.02\pm 0.06$ \\
${\cal B}(K^0_L\pi^0)$ (\%) &
$0.94\pm 0.03\pm 0.03$ & $0.96\pm 0.03\pm 0.03$ \\
${\cal B}(K^0_L\eta)$ $(10^{-3})$ &
$3.36\pm 0.30\pm 0.17$ & $3.40\pm 0.31\pm 0.17$ \\
${\cal B}(K^0_L\omega)$ (\%) &
$0.90\pm 0.05\pm 0.03$ & $0.91\pm 0.05\pm 0.03$ \\
${\cal B}(K^0_S\pi^0)$ (\%) &
$1.17\pm 0.02\pm 0.03$ & $1.16\pm 0.02\pm 0.03$ \\
${\cal B}(K^0_S\eta)$ $(10^{-3})$ &
$4.95\pm 0.14\pm 0.36$ & $4.90\pm 0.14\pm 0.36$ \\
${\cal B}(K^0_S\omega)$ (\%) &
$1.15\pm 0.02\pm 0.04$ & $1.14\pm 0.02\pm 0.04$ \\
${\cal B}(K^0_L\pi^0\pi^0)$ (\%) &
$0.95\pm 0.06\pm 0.05$ & $0.94\pm 0.06\pm 0.05$ \\
${\cal B}(K^- e^+\nu_e)$ (\%) &
$3.54\pm 0.05\pm 0.08$ & $3.52\pm 0.05\pm 0.08$ \\
${\cal B}(K^- \mu^+\nu_\mu)$ (\%) &
$3.38\pm 0.05\pm 0.08$ &  $3.36\pm 0.05\pm 0.08$ \\
${\cal B}(Y_0)$ $(10^{-3})$ &
        $4.38\pm 0.18\pm 0.12$ & $4.33\pm 0.17\pm 0.11$ \\
${\cal B}(Y_1)$ $(10^{-3})$ &
        $1.65\pm 0.10\pm 0.04$ & $1.63\pm 0.10\pm 0.04$ \\
${\cal B}(Y_2)$ $(10^{-3})$ &
        $3.43\pm 0.16\pm 0.10$ & $3.33\pm 0.14\pm 0.08$ \\
${\cal B}(Y_3)$ $(10^{-3})$ &
        $0.99\pm 0.08\pm 0.03$ & $0.97\pm 0.08\pm 0.02$ \\
${\cal B}(Y_4)$ $(10^{-3})$ &
        $1.70\pm 0.11\pm 0.05$ & $1.62\pm 0.10\pm 0.04$ \\
${\cal B}(Y_5)$ $(10^{-3})$ &
        $2.11\pm 0.13\pm 0.07$ & $2.13\pm 0.12\pm 0.05$ \\
${\cal B}(Y_6)$ $(10^{-3})$ &
        $3.15\pm 0.15\pm 0.08$ & $3.14\pm 0.14\pm 0.08$ \\
${\cal B}(Y_7)$ $(10^{-3})$ &
        $3.68\pm 0.16\pm 0.09$ & $3.71\pm 0.16\pm 0.09$ \\
\hline\hline
\end{tabular}
\end{table}

\begin{table}[htb]
\caption{Correlation coefficients (\%) for the fits in Table~\ref{tab:results1}
using both statistical and systematic uncertainties.}
\label{tab:correlations}
\begin{tabular}{lrrrr}
\hline\hline
 & ~$y$~ & ~$r^2$~ & ~$\cos\delta$~ & ~$\sin\delta$ \\
\hline
\multicolumn{5}{c}{Standard Fit} \\
$r^2$ & $0$ \\
$\cos\delta$~ & $-53$ & $-42$ \\
$\sin\delta$ & $-3$ & $+1$ & $+4$ \\
$x^2$ & $-73$ & $0$ & $+39$ & $+2$ \\
\multicolumn{5}{c}{Extended Fit} \\
$r^2$ & $+3$ \\
$\cos\delta$~ & $-27$ & $-16$ \\
$\sin\delta$ & $+62$ & $+21$ & $+36$ \\
$x^2$ & $+9$ & $+25$ & $+5$ & $-18$ \\
\hline\hline
\end{tabular}
\end{table}

These results include the first direct measurement of $\sin\delta$,
as well as first branching fraction measurements for $D^0\to K^0_L\eta$,
$K^0_L\omega$, and $K^0_L\pi^0\pi^0$. 
We also present the first CLEO-c
branching fraction measurement for $D^0\to K^-\mu^+\nu_\mu$.
In the Standard Fit,
the statistical uncertainties on $y$ and $r\cos\delta$ are approximately three
times smaller than in our previous analysis~\cite{tqca1}. 
Because of the strong
correlation between $r^2$ and $\cos\delta$, and because $r^2$ is determined by
different inputs in the two analyses, a direct comparison of the $\cos\delta$
uncertainties is not instructive.
Accounting for correlated uncertainties, the results for $\cos\delta$
in the Standard and Extended Fits differ by $2.5\sigma$. This difference is
caused primarily by the upward fluctuations in $y$ and $r^2$ in the
Standard Fit, which are both negatively correlated with $\cos\delta$.

The above factor of three improvement in $y$ and $r\cos\delta$ can be
attributed in equal parts to the increased size of the dataset and to the
additional final states in the current analysis.
In particular, the new $K^0_L$ modes (including $\{Ke\nu_e, K^0_L\pi^0\}$)
reduce the statistical uncertainties on $y$ and $r\cos\delta$ in the Standard
Fit by roughly 10\%, and the $K\mu\nu_\mu$ modes reduce them by 20--30\%.
The $K^0_S\pi^+\pi^-$ measurements from Ref.~\cite{kspipi} have a similar
effect on $y$ and $r\cos\delta$ as the new $K^0_L$ modes,
but they also provide all the information on $\sin\delta$ in the Standard Fit.

The Extended Fit demonstrates that when our $\cos\delta$ and
$\sin\delta$ measurements are used to combine $y$ and $y'$ from other
experiments, the overall uncertainty on $y$ is reduced by approximately 10\%
compared to the global average of all measurements in
Table~\ref{tab:externalYYPrime} (except $y_{CP}$ from Ref.~\cite{ycpLHCb})
found in Ref.~\cite{hfag2010}: $y = 0.79\pm 0.13$. Note that this
global average includes the results from Ref.~\cite{tqca1}.

\begin{figure}[htb]
\includegraphics*[width=\linewidth]{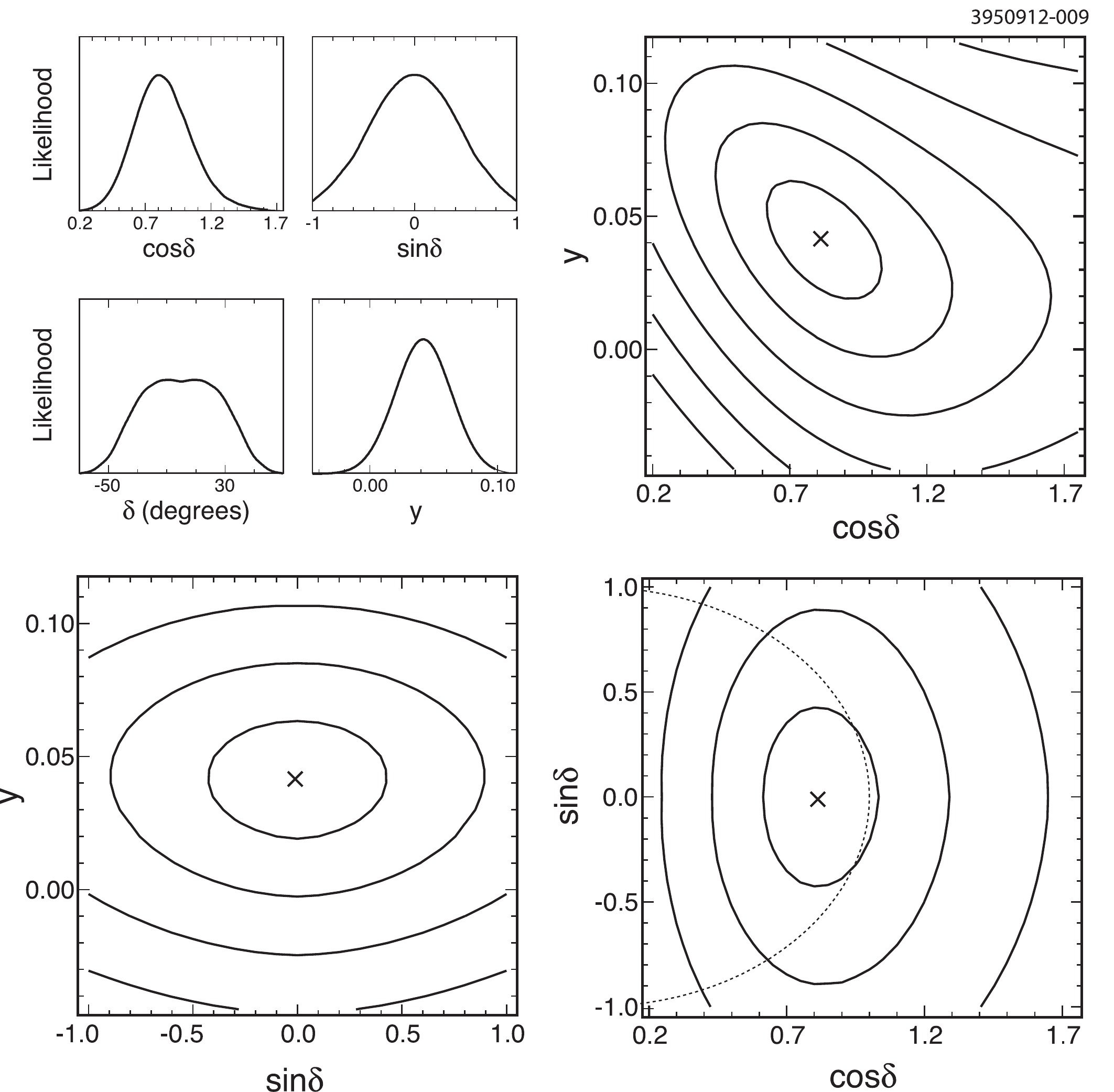}
\caption{Standard Fit likelihoods including both statistical and systematic
uncertainties for $\cos\delta$, $\sin\delta$, $\delta$, and $y$. The
two-dimensional likelihoods are shown as contours in increments of $1\sigma$,
where $\sigma=\sqrt{\Delta\chi^2}$.}
\label{fig:contoursNoExtMeas}
\end{figure}

\begin{figure}[htb]
\includegraphics*[width=\linewidth]{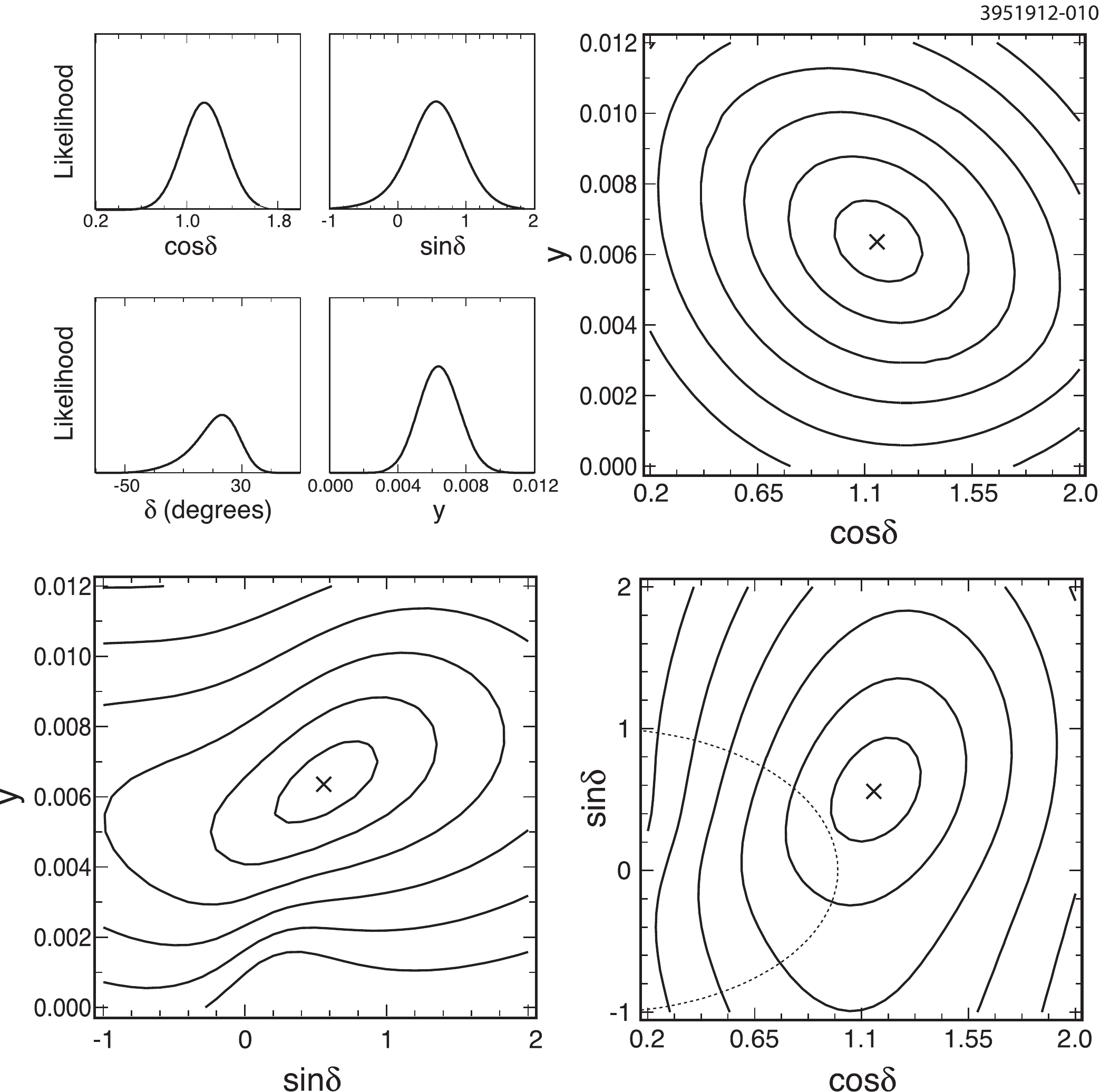}
\caption{Extended Fit likelihoods including both statistical and systematic
uncertainties for $\cos\delta$, $\sin\delta$, $\delta$, and $y$. The
two-dimensional likelihoods are shown as contours in increments of $1\sigma$,
where $\sigma=\sqrt{\Delta\chi^2}$.}
\label{fig:contoursYYPExt}
\end{figure}

Figure~\ref{fig:contoursNoExtMeas} shows the one-dimensional
posterior PDFs
for $\cos\delta$, $\sin\delta$, $\delta$, and $y$ in the
Standard Fit, including statistical and
systematic uncertainties.  These curves are
obtained by re-minimizing the $\chi^2$ at each point and computing
${\cal L} = e^{-(\chi^2-\chi^2_{\rm min})/2}$.
Also shown are the two-dimensional contours for combinations of
$y$, $\cos\delta$, and $\sin\delta$.
Because of the $\sin\delta$ sign ambiguity, the PDFs for both
$\sin\delta$ and $\delta$ in the Standard Fit are symmetric
around zero.
Figure~\ref{fig:contoursYYPExt} shows the same
distributions for the Extended Fit. Here, the sign ambiguity is
resolved by the external measurements.
All the PDFs, except those for $\delta$, are well described by
Gaussians or bifurcated Gaussians. In particular,
the non-Gaussian profile for $\cos\delta$ in the Standard Fit from
Ref.~\cite{tqca1} has been eliminated by our direct measurement
of $r^2$.

Although the central value for $\cos\delta$ in the Extended Fit
is unphysical, we find
$\tau\equiv(\cos^2\delta+\sin^2\delta)^{1/2}=1.28\pm 0.27$ to be
consistent with physical boundary. Similarly, in the Standard Fit,
$\tau=0.81\pm 0.21$.
The PDFs for $\delta$ in Figs.~\ref{fig:contoursNoExtMeas}
and~\ref{fig:contoursYYPExt}
are obtained by probing $\cos\delta$ and
$\sin\delta$ under the constraint $\tau = 1$, which reduces the
height of the PDF relative to the other parameters.
The implied values for $\delta$ from these PDFs are
$|\delta| = (10^{+28+13}_{-53-0})^\circ$ for the Standard Fit and
$\delta = (18^{+11}_{-17})^\circ$ for the Extended Fit.
Also, applying the above constraint in the Standard Fit improves
the uncertainties on $y$ and $x^2$ by 15\% and 8\%, respectively,
resulting in $y = ( 3.3\pm 1.7\pm 0.8 )\%$ and
$x^2 = (0.14\pm 0.21\pm 0.09 )\%$; the changes in all other
parameters are negligible. Performing the
Extended Fit with $\tau = 1$ produces negligible shifts
in all the fit parameters.

Our results for $c_i$, $s_i$, and branching fractions do not supersede
other CLEO-c measurements.
For $c_i$ and $s_i$, our fitted values are consistent with those in
Ref.~\cite{kspipi}, after accounting for differences between the two
analyses.

\section{Summary}\label{sec:summary}

We present an updated analysis of quantum correlations in $D^0\bar D^0$
decays at the $\psi(3770)$ using the full CLEO-c dataset, resulting in
a new value of $\cos\delta = 0.81^{+0.22+0.07}_{-0.18-0.05}$ and a
first measurement of $\sin\delta = -0.01\pm 0.41 \pm 0.04$, which, when
combined, imply a strong phase of $|\delta| = (10^{+28+13}_{-53-0})^\circ$.
By including external inputs on mixing parameters in the fit, we find
alternative values of $\cos\delta = 1.15^{+0.19+0.00}_{-0.17-0.08}$,
$\sin\delta = 0.56^{+0.32+0.21}_{-0.31-0.20}$,
and $\delta = (18^{+11}_{-17})^\circ$. The effect of these measurements
on the world averages of mixing parameters is to improve the uncertainty
on $y$ by approximately 10\%.


\acknowledgments
We gratefully acknowledge the effort of the CESR staff 
in providing us with excellent luminosity and running conditions. 
This work was supported by 
the National Science Foundation, 
the U.S. Department of Energy, 
the Natural Sciences and Engineering Research Council of Canada, and 
the U.K. Science and Technology Facilities Council. 


\end{document}